\documentclass{article}
\usepackage[top=2cm,bottom=3cm,left=2cm,right=2cm]{geometry}
\usepackage{amsmath}
\usepackage{amssymb} 
\usepackage{amsthm}
\usepackage{stmaryrd}
\usepackage{bm}
\usepackage{hyperref}
\usepackage{upgreek}
\usepackage{graphicx}
\usepackage{multirow}

\newcommand{\inverttriangle}{%
               \mathrel{\raisebox{.1em}{%
               \reflectbox{\rotatebox[origin=c]{180}{$\triangle$}}}}}

\numberwithin{equation}{section}
\numberwithin{figure}{section}

\def\eq#1{(\ref{eq:#1})}
\def\lineup{\!\!\!\!\!\!\!\!\!\!&&}

\def\d{\partial}
\def\eps{\epsilon}

\def\deg{\mathrm{deg}}

\def\M{{\bf M}}
\def\m{{\bf m}}
\def\Q{{\bf Q}}
\def\n{{\bm \upeta}}

\def\c{{\bf c}}

\def\D{{\bf D}}

\def\G{{\bf \hat{G}}}
\def\H{{\bf \hat{H}}}

\def\w{\wedge}

\def\[{\big[}
\def\]{\big]}

\def\PsiA{\Psi_A}
\def\PhiA{\Phi_A}
\def\PhiB{\Phi_B}
\def\PsiB{\Psi_B}

\begin{document}

\begin{titlepage}
\rightline{\tt LMU-ASC 63/15}
\rightline\today

\begin{center}
\vskip 3.5cm

{\large \bf{Relating Berkovits and $A_\infty$ Superstring Field Theories;\\ Large Hilbert Space Perspective}}

\vskip 1.0cm

{\large {Theodore Erler}}

\vskip 1.0cm

{\it {Arnold Sommerfeld Center, Ludwig-Maximilians University}}\\
{\it {Theresienstrasse 37, 80333 Munich, Germany}}\\
tchovi@gmail.com\\

\vskip 2.0cm

{\bf Abstract}
\end{center}

We lift the dynamical field of the $A_\infty$ superstring field theory to the large Hilbert space by introducing a gauge invariance associated with the eta zero mode. We then provide a field redefinition which relates the lifted field to the dynamical field of Berkovits' superstring field theory in the large Hilbert space. This generalizes the field redefinition in the small Hilbert space described in earlier works, and gives some understanding of the relation between the gauge symmetries of the theories. It also provides a new perspective on the algebraic structure underlying gauge invariance of the Wess-Zumino-Witten-like action.

\vskip 1.0cm
\noindent 

\noindent

\end{titlepage}

\tableofcontents

\section{Introduction}

A few recent papers \cite{OkWB,WB} have investigated the relation between the Wess-Zumino-Witten-like open superstring field theory of Berkovits \cite{Berkovits1,Berkovits2} and a new form of open superstring field theory based on $A_\infty$ algebras \cite{WittenSS}.\footnote{For corresponding investigations in heterotic string field theory \cite{BOZ,ClosedSS}, see \cite{HeteroticWB,WZWEKS}.} The first issue one encounters in this regard is that the $A_\infty$ theory uses a string field in the small Hilbert space, while the Berkovits theory uses a string field in the large Hilbert space. The large Hilbert space comes with an additional gauge symmetry associated with the eta zero mode, on top of the usual gauge symmetry associated with the BRST operator. In earlier works, this discrepancy was resolved by fixing the eta part of the gauge invariance, so that the remaining degrees of freedom of the Berkovits theory could be described by a single string field in the small Hilbert space. Then one can compare the gauge-fixed field of the Berkovits theory to the string field of the $A_\infty$ theory.
 
Here we take a complementary approach. Instead of partially gauge fixing the Berkovits theory, we lift the field of the $A_\infty$ theory to the large Hilbert space, producing a new gauge symmetry associated with the eta zero mode. This can be done as follows. One substitutes the original string field $\Psi_A$ in the $A_\infty$ action with a new dynamical field $\Phi_A$ in the large Hilbert space according to
\begin{equation}\Psi_A = \eta \Phi_A,\end{equation}
where $\eta=\eta_0$ is the eta zero mode. The lifted $A_\infty$ theory automatically possesses an additional gauge symmetry 
\begin{equation}\Phi_A' = \Phi_A +\eta \Omega_A,\end{equation}
since the action only depends on $\Phi_A$ in the combination $\eta\Phi_A$. Therefore we can search for a field redefinition relating the lifted field $\Phi_A$ to the string field $\Phi_B$ of the Berkovits theory. No gauge fixing is required. The field redefinition we propose comes in the form of a ``Wilson line" which relates a path through field space in the Berkovits theory to a path through field space in the lifted $A_\infty$ theory. When the actions are expressed in Wess-Zumino-Witten-like form, this reduces their equivalence to an identity.

One advantage of this approach is that it gives a clearer understanding of the relation between the gauge symmetries of the two theories. This question is more difficult to study in the small Hilbert space since the partially gauge-fixed Berkovits theory does not exhibit a cyclic $A_\infty$ structure, and the full nonlinear gauge invariance has not been completely worked out.\footnote{The nonlinear gauge invariance of the partially gauge-fixed Berkovits theory has been worked out to second order in \cite{INOT}.} We show that the $\eta$ gauge transformations of the two theories map into each other, while the BRST gauge transformation in one theory maps into a combination of BRST, $\eta$, and trivial gauge transformations in the other. Another consequence of our analysis is a new perspective on the algebraic structure underlying the Wess-Zumino-Witten-like (WZW-like) action. We show that the ``potentials" which appear in the WZW-like action are generally part of a hierarchy of higher-form potentials which together provide a solution to a certain Maurer-Cartan equation. Maurer-Cartan gauge transformations implement field redefinitions and relate equivalent realizations of the WZW-like action. In this sense, the Maurer-Cartan equation plays a role in the large Hilbert space somewhat analogous to the role of cyclic $A_\infty$ algebras in the small Hilbert space. These results may be useful step towards a better understanding of the role of the large Hilbert space in superstring field theory, and in particular the problem of quantization \cite{BigBV,BerkovitsBV,Torii1,Torii2,Torii3}.

\section{Recap}

In this section contains a repository of definitions and formulae that we will need in our calculations. See earlier works for a more extended introduction to the formalism, especially \cite{WB}, whose conventions we follow. 

In this paper, string fields are always elements of the Neveu-Schwarz state space $\mathcal{H}$ of an open superstring quantized in the RNS formalism, including $bc$ and bosonized superconformal ghosts $\eta,\xi,e^{\phi}$ \cite{FMS}. A string field $A$ is in the small Hilbert space if $\eta A=0$, otherwise it is in the large Hilbert space. The {\it degree} of a string field $A$, denoted $\deg(A)$, is defined to be its Grassmann parity $\eps(A)$ plus one:
\begin{equation}\deg(A) = \eps(A)+1\ \ \mathrm{mod}\ \mathbb{Z}_2.\end{equation}
 When using the degree grading, it is natural to work with a 2-product $m_2$ and a symplectic form $\omega_L$ related, respectively, to Witten's open string star product and the BPZ inner product by a sign:
\begin{eqnarray}
m_2(A,B)\lineup = (-1)^{\deg(A)}A*B,\\
\omega_L(A,B)\lineup = (-1)^{\deg(A)}\langle A,B\rangle_L.
\end{eqnarray}
We will often drop the star when writing the star product, that is  $AB\equiv A*B$. The subscript $L$ denotes the BPZ inner product and symplectic form computed in the large Hilbert space. We will also encounter the symplectic form in the small Hilbert space denoted $\omega_S(A,B)$.  We will omit the subscript $S$ or $L$ for equations that hold for symplectic forms in both the large and small Hilbert space. The definition of the $A_\infty$ theory requires an operator built from the $\xi$ ghost
\begin{equation}\xi \equiv \oint_{|z|=1}\frac{dz}{2\pi i} f(z)\xi(z),\label{eq:xi}\end{equation}
where $f(z)$ is a function which is holomorphic in the neighborhood of the unit circle, $\xi$ is BPZ even, and $[\eta,\xi]=1$.\footnote{In contrast to \cite{WB}, in this paper we set the open string coupling constant to $1$.} Using this operator, the symplectic form in the small Hilbert space can be related to the symplectic form in the large Hilbert space by
\begin{equation}\omega_S(A,B) = \omega_L(\xi A,B),\end{equation}
where $A$ and $B$ are string fields in the small Hilbert space. 

An $n$-string product $c_n(A_1,...,A_n)$ can be viewed as a linear map from $n$ copies of the state space $\mathcal{H}$ into one copy:
\begin{equation}c_n:\mathcal{H}^{\otimes n}\to\mathcal{H}.\end{equation}
We write
\begin{equation}c_n(A_1,...,A_n)=c_n \, A_1\otimes ... \otimes A_n  ,\end{equation}
where the right hand side is interpreted as the linear map $c_n$ acting on the tensor product of states $A_1\otimes ... \otimes A_n$. The degree of $c_n$ is defined to be the degree of its output minus the sum of the degrees of its inputs. We consider the tensor algebra $T\mathcal{H}$ generated by taking sums of tensor products of states:
\begin{equation}T\mathcal{H} = \mathcal{H}^{\otimes 0}\, \oplus\,\mathcal{H}\, \oplus\, \mathcal{H}^{\otimes 2}\, \oplus\, ...\ .\end{equation}
Here $\mathcal{H}^{\otimes 0}$ consists of scalar multiples of the identity element of the tensor algebra $1_{T\mathcal{H}}$, satisfying 
\begin{equation}1_{T\mathcal{H}}\otimes A =  A\otimes 1_{T\mathcal{H}} = A,\end{equation}
for any $A\in T\mathcal{H}$. Suppose we have a list consisting of a 0-string product $D_0$, a 1-string product $D_1$, a 2-string product $D_2$ and so on so that all products are either degree even or degree odd. From this data we define an operator on the tensor algebra called a {\it coderivation}
\begin{equation}\D = \sum_{\ell,m,n=0}^{\infty} (\mathbb{I}^{\otimes \ell} \otimes D_m \otimes \mathbb{I}^{\otimes n})\pi_{\ell+m+n}.\end{equation}
Here $\pi_n$ denotes the projection onto the $n$-string component of the tensor algebra and $\mathbb{I}^{\otimes n}$ is the identity operator on 
$\mathcal{H}^{\otimes n}$. The tensor product of linear maps acts on tensor products of states as follows. If $b_{k,m}$ is a linear map $\mathcal{H}^{\otimes m}\to\mathcal{H}^{\otimes k}$ and $c_{\ell,n}$ is a linear map $\mathcal{H}^{\otimes \ell}\to\mathcal{H}^{\otimes n}$, then their tensor product is defined to acts as  
\begin{equation}(b_{k,m}\otimes c_{\ell,n}) A_1\otimes ...\otimes A_{m+n} = (-1)^{\deg(c_{\ell,n})(\deg(A_1) + ... +\deg(A_m))}
(b_{k,m}\, A_1\otimes ... \otimes A_m)\otimes (c_{\ell,n}\, A_{m+1}\otimes ... \otimes A_{m+n}).
\end{equation}
A product $c_n$ defines a coderivation $\c_n$ constructed by taking $D_n = c_n$ and $D_{m\neq n} = 0$. Now 
suppose we have a list of degree even products $H_0, H_1,H_2,...$. With this data we can define an operator on the tensor algebra called a {\it cohomomorphism}
\begin{equation}\H = \pi_0 + \sum_{\ell=1}^{\infty} \sum_{k_1,...,k_\ell=0}^{\infty} (H_{k_1}\otimes ... \otimes H_{k_\ell})\pi_{k_1...+k_\ell}.\end{equation}
A typical example of a cohomomorphism is the identity operator on the tensor algebra $\mathbb{I}_{T\mathcal{H}}$, which is defined by taking $H_1=\mathbb{I}$ and the remaining $H_k$ to vanish. Given a degree even string field $A$, we can define an element of the tensor algebra called a {\it group-like element}:
\begin{equation}\frac{1}{1-A}\equiv 1_{T\mathcal{H}} + A + (A\otimes A) + (A\otimes A\otimes A)+...\ .\end{equation}
In our calculations we will often need to act coderivations and cohomomorphisms on group-like elements. We note the formulas 
\begin{eqnarray}
\D\frac{1}{1-A} \lineup = \frac{1}{1-A}\otimes\left(\pi_1\D\frac{1}{1-A}\right)\otimes \frac{1}{1-A},\label{eq:Dgrp}\\
\H\frac{1}{1-A}\lineup = \frac{1}{1-\pi_1\H\frac{1}{1-A}}.\label{eq:Hgroup}
\end{eqnarray}
By taking variations we can obtain several useful generalizations. For example, if we define the variation $\delta A = B$, then taking the variation of both sides of \eq{Dgrp} gives
\begin{eqnarray}
\D\frac{1}{1-A}\otimes B\otimes\frac{1}{1-A} \lineup = \frac{1}{1-A}\otimes\left(\pi_1\D\frac{1}{1-A}\right)\otimes \frac{1}{1-A}\otimes B\otimes
\frac{1}{1-A}\nonumber\\
\lineup\ \ \ +
\frac{1}{1-A}\otimes\left(\pi_1\D \frac{1}{1-A}\otimes B\otimes \frac{1}{1-A}\right)\otimes \frac{1}{1-A} \nonumber\\
\lineup\ \ \ + (-1)^{\deg(\D)\deg(B)}\frac{1}{1-A}\otimes B\otimes \frac{1}{1-A}\otimes\left(\pi_1\D\frac{1}{1-A}\right)\otimes \frac{1}{1-A}.
\end{eqnarray}
Other formulas can be derived similarly. A {\it symplecitc form} $\omega$ is a linear map from two copies of the state space into complex numbers:
\begin{equation}\langle \omega|: \mathcal{H}^{\otimes 2}\to\mathbb{C}.\end{equation}
which is graded antisymmetric upon interchange of its arguments:
\begin{equation}\omega(A,B) = -(-1)^{\deg(A)\deg(B)}\omega(B,A).\end{equation}
We write $\langle \omega| A\otimes B = \omega(A,B)$. An $n$-string product $c_n$ is {\it cyclic} with respect to the symplectic form $\omega$ if its coderivation ${\bf c}_n$ satisfies 
\begin{equation}\langle \omega| \pi_2 \c_n = 0.\end{equation}
Likewise, a cohomomorphism $\H$ is {\it cyclic} if it satisfies 
\begin{equation}\langle\omega|\pi_2\H = \langle \omega| \pi_2.\end{equation}
This summarizes most of what we need from the coalgebra formalism. At the margins, a few computations are helped by introducing the {\it coproduct}. We will review this in appendix \ref{app:cyclic}.

Now let's review the $A_\infty$ superstring field theory. The dynamical string field $\PsiA$ is in the Neveu-Schwarz (NS) sector,\footnote{In this paper we only discuss the NS sector. The generalization to the Ramond sector \cite{Ramond,OKish,SenR1,SenR2,MatsunagaR} will be considered in \cite{Ok2}.} is degree even, lives in the small Hilbert space, and carries ghost number 1 
and picture number $-1$. The action is 
\begin{equation}S_{A} = \frac{1}{2}\omega_S(\PsiA,Q\PsiA) +\frac{1}{3}\omega_S(\PsiA,M_2(\PsiA,\PsiA)) + \frac{1}{4}\omega_S(\PsiA,M_3(\PsiA,\PsiA,\PsiA))+...,\end{equation}
where $Q\equiv Q_B$ is the BRST operator and $Q,M_2,M_3,...$ are a sequence of degree odd multi-string products in the small Hilbert space which satisfy the relations of a cyclic $A_\infty$ algebra. The products $Q,M_2,M_3,...$ define a coderivation 
\begin{equation}\M = \Q+ \M_2+\M_3+...\ .\end{equation}
The statement that the products are in the small Hilbert space can be expressed by the equation\footnote{Commutators of products and coderivations are always graded with respect to degree. Commutators of string fields, with the multiplication defined by Witten's open string star product, are always graded with respect to Grassmann parity.} 
\begin{equation}[\n,\M] = 0,\end{equation}
where $\n$ is the coderivation corresponding to the eta zero mode. The statement that the products form an $A_\infty$ algebra is expressed by the equation 
\begin{equation}\M^2 = 0.\end{equation}
In addition, the products form a {\it cyclic} $A_\infty$ algebra because 
\begin{equation}\langle \omega| \pi_2 \M = 0.\end{equation}
A key property of the $A_\infty$ superstring field theory is that the coderivation $\M$ can be related to $\Q$ using a similarity transformation in the large Hilbert space. The similarity transformation is provided by an invertible, cyclic cohomomorphism $\G$ 
\begin{equation}\langle \omega_L|\pi_2\G = \langle \omega_L|\pi_2\end{equation}
satisfying \cite{WittenSS,OkWB}
\begin{eqnarray}\M = \G^{-1}\Q\G,\ \ \ \ \n -\m_2 = \G\n \G^{-1},\label{eq:GQn}\end{eqnarray}
where $\m_2$ is the coderivation corresponding to the open string star product $m_2$. The construction of $\G$ requires the operator $\xi$ in \eq{xi}, and is described in \cite{WittenSS}. 

\section{$A_\infty$ Action in the Large Hilbert Space}

In this section we reformulate the $A_\infty$ superstring field theory by replacing $\PsiA$ in the small Hilbert space with a new dynamical field $\PhiA$ in the large Hilbert space via the substitution
\begin{equation}\PsiA = \eta\PhiA.\label{eq:Asml}\end{equation}
We then reexpress the action in Wess-Zumino-Witten-like form. The key aspects of the derivation follow \cite{OkWB}, but we will make some refinements. 

The $n$-string vertex in the $A_\infty$ action takes the form
\begin{equation}\frac{1}{n}\omega_S(\PsiA,M_{n-1}(\PsiA,...,\PsiA)).\end{equation}
It will be convenient to eliminate the multiplicative factor of $1/n$. This can be done with the following trick. Introduce a 1-parameter family of string fields $\PsiA(t),t\in[0,1]$ satisfying the boundary conditions 
\begin{equation}\Psi_A(0) = 0;\ \ \  \PsiA(1)=\PsiA.\end{equation}
We refer to $\PsiA(t)$ as an {\it interpolating field}, or simply {\it interpolation}. We can write the action as an integral of a total derivative with respect to $t$:
\begin{equation}S_{A} = \int_0^1dt\,\frac{d}{dt}\left[\frac{1}{2}\omega_S\big(\PsiA(t),Q\PsiA(t)\big)+ \frac{1}{3}\omega_S\big(\PsiA(t),M_2(\PsiA(t),\PsiA(t))\big) 
+ ...\right].\end{equation}
Acting the $t$-derivative on the $n$-string vertex produces $n$ terms containing a factor of $\dot{\Psi}_A(t) = d\PsiA(t)/dt$. Since the vertices are cyclic, each of these terms are equal, canceling the factor of $1/n$. Using cyclicity to place $\dot{\Psi}_A(t)$ in the first entry of the symplectic form, we can therefore write the action 
\begin{equation}S_{A} =\int_0^1dt\left[\omega_S\big(\dot{\Psi}_A(t),Q\PsiA(t)\big)+ \omega_S\big(\dot{\Psi}_A(t),M_2(\PsiA(t),\PsiA(t))\big) 
+...\right],\end{equation}
which can be written more compactly as 
\begin{equation}S_{A} = \int_0^1dt\,\omega_S\left(\dot{\Psi}_A(t),\pi_1\M \frac{1}{1-\PsiA(t)}\right).\end{equation}
Since this form of the action was obtained from the integral of a total derivative, by construction it only depends on the value of $\PsiA(t)$ at $t=1$.

The next step is to lift to the large Hilbert space by making the substitution
\begin{equation}\PsiA(t) = \eta\PhiA(t),\end{equation}
where $\PhiA(t)$ is an interpolating 1-parameter family of string fields subject to the boundary conditions
\begin{equation}\Phi_A(0) = 0;\ \ \  \PhiA(1)=\PhiA,\end{equation}
and $\PhiA$ is the new dynamical string field in the large Hilbert space. The new field $\PhiA$ is degree odd (but Grassmann even) and carries ghost and picture number zero. The action becomes 
\begin{equation}S_{A} = \int_0^1dt\,\omega_L\left(\xi \eta \dot{\Phi}_A(t),\pi_1\M \frac{1}{1-\eta\PhiA(t)}\right),\label{eq:action3}\end{equation}
where we replaced the small Hilbert space symplectic form by the large Hilbert space symplectic form. The action only depends on the value of $\PhiA(t)$ at $t=1$ (in fact, it only depends on the value of $\eta\PhiA(t)$ at $t=1$), and has a new gauge invariance related to the eta zero mode
\begin{equation}\delta_\eta \Phi_A = \eta \Omega_A,\label{eq:deltaeta}\end{equation}
where the gauge parameter $\Omega_A$ is degree even, ghost number $-1$ and picture number $1$. The infinitesimal BRST gauge transformation is
\begin{equation}\delta_Q \PhiA = \pi_1\M \frac{1}{1-\eta\PhiA}\otimes \Lambda_A\otimes \frac{1}{1-\eta\PhiA},\label{eq:deltaQ}\end{equation}
where the gauge parameter $\Lambda_A$ is degree even, ghost number $-1$ and picture 0. Acting $\eta$ on both sides of this equation produces the standard $A_\infty$ gauge transformation of $\PsiA=\eta\PhiA$, with gauge parameter $-\eta\Lambda_A$. Note that in the action \eq{action3}, the second entry of the symplectic form is in the small Hilbert space. Therefore in the first entry of the symplectic form we can replace $\xi\eta$ with $\xi\eta+\eta\xi = 1$ at no cost. We can therefore write
\begin{equation}S_{A} = \int_0^1dt\,\omega_L\left(\dot{\Phi}_A(t),\pi_1\M \frac{1}{1-\eta\PhiA(t)}\right).\label{eq:action4}\end{equation}
To simplify further, recall from \cite{OkWB} that a cyclic cohomomorphism $\H$ satisfies the identity
\begin{equation}\omega\left(\pi_1\D_1\frac{1}{1-A},\pi_1 \D_2\frac{1}{1-A}\right)=\omega\left(\pi_1\H\D_1\frac{1}{1-A},\pi_1 \H \D_2\frac{1}{1-A}\right),\label{eq:Id1}\end{equation}
where $A$ is a string field and $\D_1$ and $\D_2$ are arbitrary coderivations. We provide another proof of this identity in appendix \ref{app:cyclic}. Since the cohomomorphism $\G$ is cyclic with respect to $\omega_L$, this identity allows us to write 
\begin{equation}\omega_L\left(\dot{\Phi}_A(t),\pi_1\M \frac{1}{1-\eta\PhiA(t)}\right) = \omega_L\left(\pi_1\G\dot{\bf{\Phi}}_A(t)\frac{1}{1-\eta\PhiA(t)},\pi_1\G \M \frac{1}{1-\eta\PhiA(t)}\right),\end{equation}
where $\dot{\bf{\Phi}}_A(t)$ is the coderivation corresponding to the string field $\dot{\Phi}_A(t)$ regarded as a zero-string product. We can further write
\begin{eqnarray}
\pi_1\G\dot{\bf{\Phi}}_A(t)\frac{1}{1-\eta\PhiA(t)} \lineup = \pi_1\G\frac{1}{1-\eta\PhiA(t)}\otimes\dot{\Phi}_A(t)\otimes\frac{1}{1-\eta\PhiA(t)},\\
\pi_1\G \M \frac{1}{1-\eta\PhiA(t)} \lineup = \pi_1\Q\G \frac{1}{1-\eta\PhiA(t)}.
\end{eqnarray}
The action \eq{action4} therefore becomes
\begin{equation}
S_{A} = \int_0^1dt\,\omega_L\left(\pi_1\G\frac{1}{1-\eta\PhiA(t)}\otimes\dot{\Phi}_A(t)\otimes\frac{1}{1-\eta\PhiA(t)},Q
\pi_1\G \frac{1}{1-\eta\PhiA(t)}\right).
\end{equation}
Now we define ``potentials"
\begin{eqnarray}
A_t \lineup \equiv \pi_1\G\frac{1}{1-\eta\PhiA(t)}\otimes\dot{\Phi}_A(t)\otimes\frac{1}{1-\eta\PhiA(t)},\\
A_\eta \lineup \equiv \pi_1\G \frac{1}{1-\eta\PhiA(t)}.
\end{eqnarray}
The potential $A_t$ is degree odd but Grassmann even. The potential $A_\eta$ is degree even but Grassmann odd. Switching to the Grassmann grading the action is therefore expressed
\begin{equation}S_A = -\int_0^1 dt\,\langle A_t,QA_\eta\rangle_L.\label{eq:WZW}\end{equation}
This is the Wess-Zumino-Witten-like form of the action for the lifted $A_\infty$ theory. Note that with this definition of $A_t$ and $A_\eta$ we will not be able to express the action in the standard WZW-like form
\begin{equation}
S_A \neq -\frac{1}{2}\langle A_\eta,A_Q\rangle_L\Big|_{t=1} - \frac{1}{2}\int_0^1 dt\, \langle A_t,[A_\eta,A_Q]\rangle_L.\label{eq:stWZW}
\end{equation}
A similar obstruction prevents the action of heterotic string field theory from being expressed in standard WZW-like form. The action \eq{WZW} turns out to be more general, and will be the basis of our computations.

The above embedding of the $A_\infty$ theory into the large Hilbert space is trivial---the action only depends on the large Hilbert space field in the combination $\eta\PhiA$. One might ask whether there is a  more interesting extension of the $A_\infty$ theory to the large Hilbert space. In this regard it is useful compare to the partially gauge-fixed Berkovits theory \cite{INOT,WB}, whose extension to the large Hilbert space should apparently be Berkovits' open superstring field theory. If $\PsiB$ is the field of the partially gauge-fixed Berkovits theory, we can attempt to replace it with a field $\PhiB$ in the large Hilbert space in the same way as we did for the $A_\infty$ theory: 
\begin{equation}\PsiB = \eta\PhiB\ ?\end{equation}
However, the resulting theory in the large Hilbert space is not Berkovits' superstring field theory. To get the Berkovits theory we must do something more refined. First we start with the potentials expressed in terms of the star product, $\xi$, and $\PsiB$. Then, for every instance where $\xi$ operates on $\PsiB$, we should substitute $\xi\PsiB$ with $\PhiB$, and for every other case we should substitute $\PsiB$ with $\eta\PhiB$. The resulting potentials define the WZW-like action for Berkovits' superstring field theory. In principle, we can follow the same procedure for the $A_\infty$ theory, producing a string field theory in the large Hilbert space with a nonlinear $\eta$ gauge invariance. However, for the Berkovits theory this procedure has a special property: it allows one to eliminate all instances of $\xi$ from the partially gauge-fixed action, so that the theory in the large Hilbert space can be expressed solely in terms of $Q,\eta$ and the star product. In the $A_\infty$ theory the same procedure does not eliminate all insertions of $\xi$, since in some cases $\xi$ acts on products of fields rather than the field itself. The upshot is that we do not know of a nontrivial embedding of the $A_\infty$ in the large Hilbert space which is more ``natural" than the trivial one. The trivial embedding is sufficient for our purposes, and will be the focus for the remainder of the paper.

\subsection{Potentials and Field Strengths}
\label{subsec:PF}
 
The structure of the WZW-like action can be understood in terms of the properties of potentials and field strengths. Suppose that we have a collection of graded derivations of the open string star product which commute. We write the derivations $\d_I$ with $I$ ranging over an index set, and they satisfy
\begin{equation}[\d_I,\d_J] =0,\label{eq:dcomm}\end{equation}
where the commutator $[,]$ is graded with respect to Grassmann parity. In particular, if $\d_I$ is Grassmann odd we have $\d_I^2=0$. Suppose that for each $\d_I$ we have an associated string field $A_I$ with the same Grassmann parity, ghost and picture number. We refer to $A_I$ as the {\it potential} corresponding to the derivation $\d_I$. 
The {\it field strength} is defined 
\begin{equation}F_{IJ} \equiv \d_I A_{J} - (-1)^{\eps(I)\eps(J)} \d_J A_{I}-[A_{I},A_{J}],\label{eq:F}\end{equation} 
where $\eps(I)$ denotes the Grassmann parity of $\d_I$, and the commutator of string fields is computed with the star product and is graded with respect to Grassmann parity. The field strength is graded antisymmetric:
\begin{equation}F_{IJ} = -(-1)^{\eps(I)\eps(J)}F_{JI}.\end{equation}
Note that diagonal elements of the field strength do not necessarily vanish. If the derivation $\d_I$ is Grassmann odd, we have
\begin{equation}F_{II} = 2(\d_I A_I -A_I*A_I),\ \ \ \eps(I) = 1\ \mathrm{mod}\ \mathbb{Z}_2.\end{equation}
It is useful to define a gauge covariant derivative
\begin{equation}\nabla_I \Psi \equiv \d_I\Psi - [A_I,\Psi].\end{equation}
We have the identities 
\begin{equation}\d_I A_{J} = (-1)^{\eps(I)\eps(J)}\nabla_{J} A_{I}+F_{IJ}\label{eq:dswitch}\end{equation}
and
\begin{equation}[\nabla_{I},\nabla_{J}]\Psi =- [F_{IJ},\Psi].\end{equation}
In particular, covariant derivatives commute if the associated field strength vanishes. 

When computing the variation of the WZW-like action we need four derivations of the star product, with associated potentials:
\begin{eqnarray}
\eta\lineup\ \leftrightarrow\ A_\eta,\ \ \ \ \ \eps(\eta)=1,\\
d/dt\lineup\ \leftrightarrow\ A_t,\ \ \ \ \ \eps(d/dt)=0,\\
\delta\lineup\ \leftrightarrow\ A_\delta,\ \ \ \ \ \eps(\delta)=0,\\
Q\lineup\ \leftrightarrow\ A_Q,\ \ \ \ \ \eps(Q)=1.
\end{eqnarray}
The variational derivative $\delta$ denotes an arbitrary variation of the interpolating field $\PhiA(t)$. In the lifted $A_\infty$ theory, the potentials are naturally defined by
\begin{eqnarray}
A_\eta(t) \lineup \equiv \pi_1\G\frac{1}{1-\eta\PhiA(t)},\label{eq:A0}\\
A_t(t)\lineup \equiv \pi_1\G\frac{1}{1-\eta\PhiA(t)}\otimes\dot{\Phi}_A(t)\otimes\frac{1}{1-\eta\PhiA(t)},\label{eq:At}\\
A_\delta(t)\lineup \equiv \pi_1\G \frac{1}{1-\eta\PhiA(t)}\otimes\delta\PhiA(t)\otimes\frac{1}{1-\eta\PhiA(t)},\label{eq:Adelta}\\
A_Q(t) \lineup \equiv \pi_1 \G\frac{1}{1-\eta\PhiA(t)}\otimes a_Q(t) \otimes\frac{1}{1-\eta\PhiA(t)},\label{eq:AQ}
\end{eqnarray}
where in the last equation the string field $a_Q(t)$ is defined
\begin{equation}
a_Q(t) \equiv \int_0^t ds\, \pi_1\M\frac{1}{1-\eta\PhiA(s)}\otimes\dot{\Phi}_A(s)\otimes\frac{1}{1-\eta\PhiA(s)}.\label{eq:aQ}
\end{equation}
A comment about notation: The potentials, field strengths, and related objects are functions of the interpolating variable $t$. To avoid clutter, we will often leave this implicit. When the dependence is not explicitly indicated, we will always assume that the interpolating variable has been set equal to $t$. So, for example, $A_\delta$ should be interpreted as $A_\delta(t)$, and $\nabla_\eta$ should be interpreted as $\nabla_\eta(t)$. The exception to this rule will be the dynamical string field and gauge parameters, where the dependence on $t$ is always indicated except when $t=1$.

The key property of the potentials \eq{A0}-\eq{AQ} is that the associated field strengths vanish along the $\eta$ direction:
\begin{eqnarray}
F_{\eta\eta}\lineup =0,\\
F_{t\eta}\lineup =0,\\
F_{\delta\eta}\lineup =0,\\
F_{Q\eta}\lineup =0.
\end{eqnarray}
As we will review in a moment, this property implies the expected formula for the variation of the WZW-like action. Therefore, the vanishing of these field strengths is the basis for the claim that the lifted $A_\infty$ action can be written in WZW-like form. It is not difficult to show that these field strengths vanish by direct substitution of the provided definitions \cite{OkWB}, but it will be helpful to give a slightly more abstract argument. Suppose the list derivations $\d_I$ includes $\eta$ and other derivations which we denote collectively as $\d_i$ (with a lower case index):
\begin{equation}\d_I \ \leftrightarrow \ \eta\, ,\, \d_i .\end{equation}
 For example, $\d_i$ should include $Q,d/dt$ and $\delta$. We assume that the $\d_i$s commute among themselves and with $\eta$. For each $\d_i$ there is a natural associated potential in the lifted $A_\infty$ theory: 
\begin{equation}
A_i(t) \equiv \pi_1\G\frac{1}{1-\eta\PhiA(t)}\otimes a_i(t)\otimes\frac{1}{1-\eta\PhiA(t)},\label{eq:Ad}
\end{equation}
The string field $a_i(t)$ will be called the {\it little potential}, and is defined 
\begin{equation}a_i(t)\equiv \int_0^t ds\, \pi_1\D_i\frac{1}{1-\eta\PhiA(s)}\otimes\dot{\Phi}_A(s)\otimes\frac{1}{1-\eta\PhiA(s)},\label{eq:ad}\end{equation}
where the coderivation $\D_i$ is the image of ${\bm \partial}_i$ under mapping with $\G$: 
\begin{equation}{\bm \partial}_i \G = \G\D_i,\end{equation}
and ${\bm \partial}_i$ is the coderivation corresponding to $\d_i$. This definition agrees with $a_Q(t)$ given in \eq{aQ}, while in the $t$ and $\delta$ directions it simplifies to 
\begin{equation}a_t(t) = \dot{\Phi}_A(t),\ \ \ a_\delta(t) = \delta\PhiA(t),\end{equation}
since $d/dt$ and $\delta$ happen to commute with $\G$. The little potential satisfies the identity
\begin{equation}\eta a_i(t) + (-1)^{\deg(i)+1}\D_i\frac{1}{1-\eta\PhiA(t)}=0,\label{eq:little1}\end{equation}
where $\deg(i) = \eps(i)$ is the degree of $\d_i$. To see that this identity holds, note that $\n$ commutes with the $\D_i$s:
\begin{equation}[\n,\D_i] = \G^{-1}[\n-\m_2,{\bm \partial_i}]\G = 0,\end{equation}
since the $\d_i$s commute with $\eta$ and are derivations of the open string star product. By a similar computation one can show that $[\D_i,\D_j]=0$. Now act $\eta$ on the little potential $a_i$ and compute: 
\begin{eqnarray}
\eta a_i(t) \lineup = (-1)^{\deg(i)}\int_0^t ds\, \pi_1\D_i \frac{1}{1-\eta\PhiA(s)}\otimes\eta\dot{\Phi}_A(s)\otimes\frac{1}{1-\eta\PhiA(s)},\\
\lineup = (-1)^{\deg(i)}\int_0^t ds\frac{d}{ds} \pi_1\D_i\frac{1}{1-\eta\PhiA(s)},\\
\lineup = (-1)^{\deg(i)}\pi_1\D_i\frac{1}{1-\eta\PhiA(t)}.
\end{eqnarray}
Now let us demonstrate that the field strength $F_{i\eta}$ vanishes. Compute 
\begin{eqnarray}
\nabla_\eta A_i \lineup = \pi_1(\n -\m_2)\G\frac{1}{1-\eta\PhiA(t)}\otimes a_i(t) \otimes\frac{1}{1-\eta\PhiA(t)},\\
\lineup = \pi_1\G\n\frac{1}{1-\eta\PhiA(t)}\otimes a_i(t) \otimes\frac{1}{1-\eta\PhiA(t)},\\
\lineup = \pi_1\G\frac{1}{1-\eta\PhiA(t)}\otimes \eta a_i(t) \otimes\frac{1}{1-\eta\PhiA(t)}.
\end{eqnarray}
Plugging in \eq{little1}, 
\begin{eqnarray}
\nabla_\eta A_i \lineup = (-1)^{\deg(i)}\pi_1\G\frac{1}{1-\eta\PhiA(t)}\otimes \left(\pi_1\D_i\frac{1}{1-\eta\PhiA(t)}\right) \otimes\frac{1}{1-\eta\PhiA(t)},\\
\lineup = (-1)^{\deg(i)}\pi_1\G\D_i\frac{1}{1-\eta\PhiA(t)},\\
\lineup = (-1)^{\deg(i)}\d_i\pi_1\G\frac{1}{1-\eta\PhiA(t)},\\
\lineup = (-1)^{\deg(i)}\d_i A_\eta.
\end{eqnarray}
Therefore, the field strength $F_{i\eta} =  \d_i A_\eta-(-1)^{\eps(i)}\nabla_\eta A_i$ vanishes as claimed. The proof that $F_{\eta\eta}$ vanishes is a different but straightforward computation, given in \cite{OkWB}.

Berkovits' open superstring field theory is defined by a dynamical field $\PhiB$ in the large Hilbert space with the same quantum numbers as $\PhiA$. The action is defined by the potentials
\begin{eqnarray}B_\eta(t)\lineup \equiv (\eta e^{\PhiB(t)})e^{-\PhiB(t)},\ \ \ B_t(t)\equiv \left(\frac{d}{dt} e^{\PhiB(t)}\right)e^{-\PhiB(t)},\nonumber\\
B_\delta(t)\lineup \equiv (\delta e^{\PhiB(t)})e^{-\PhiB(t)},\ \ \ B_Q(t)\equiv (Q e^{\PhiB(t)})e^{-\PhiB(t)}.\label{eq:Bpotentials}
\end{eqnarray}
For a general derivation $\d_i$ the potential is
\begin{equation}B_i(t)\equiv (\d_i e^{\PhiB(t)})e^{-\PhiB(t)}.\end{equation}
All field strengths in the Berkovits theory vanish. By contrast, except along the $\eta$ direction, field strengths in the lifted $A_\infty$ theory do not vanish.  However, they satisfy the Bianchi identity:
\begin{equation}\nabla_{I}F_{JK}+(-1)^{\eps(I)(\eps(J)+\eps(K))}\nabla_{J}F_{KI}+(-1)^{\eps(K)(\eps(I)+\eps(J))}\nabla_{K}F_{IJ} = 0.\end{equation} 
Suppose we choose $\d_I$ to be $\eta$. Then the second two terms in this equation vanish, and we find that the nonvanishing field strengths are covariantly constant in the $\eta$ direction:
\begin{equation}\nabla_\eta F_{t\delta} = 0,\ \ \ \nabla_\eta F_{tQ} = 0,\ \ \ \nabla_\eta F_{\delta Q} = 0,\ \ \ \nabla_\eta F_{QQ} = 0.\end{equation}
We will have more to say about the nonvanishing field strengths later, but for now we have all the ingredients needed to compute the variation of the WZW-like action and establish gauge invariance. 

\subsection{Variation of the Action and Gauge Invariance}
\label{subsec:var}

Consider the variation of the integrand of the WZW-like action:
\begin{eqnarray}\delta \langle A_t,QA_\eta\rangle_L \lineup = \langle \delta A_t,QA_\eta\rangle_L+\langle A_t,Q\delta A_{\eta}\rangle_L,\nonumber\\
\lineup = \langle \nabla_t A_\delta,QA_\eta\rangle_L+\langle F_{\delta t},QA_\eta\rangle_L +\langle Q\nabla_\eta A_{\delta},A_t\rangle_L,\label{eq:var1}
\end{eqnarray}
where in the second step we applied \eq{dswitch} to express the variation in terms of $A_\delta$. The term with the field strength actually vanishes because
\begin{equation}\langle F_{\delta t},QA_\eta\rangle_L = -\langle F_{\delta t},\nabla_\eta A_Q\rangle_L ,\end{equation}
and the field strength is annihilated by $\nabla_\eta$. Pulling $Q$ and the $\nabla_\eta$ off the $A_\delta$ in the last term of \eq{var1}, we therefore have
\begin{equation}\delta \langle A_t,QA_\eta\rangle_L = \langle \nabla_t A_\delta,QA_\eta\rangle_L- \langle A_{\delta},\nabla_\eta QA_t\rangle_L.\end{equation}
Now apply \eq{dswitch} again in the second term to replace $Q$ by a $t$ derivative:
\begin{equation}\delta \langle A_t,QA_\eta\rangle_L = \langle \nabla_t A_\delta,QA_\eta\rangle_L- \langle A_{\delta},\nabla_\eta (\nabla_tA_Q + F_{Qt})\rangle_L.\end{equation}
Again the field strength does not contribute since it is annihilated by $\nabla_\eta$. Moreover, we can commute $\nabla_\eta$ and $\nabla_t$ since $F_{t\eta}=0$.
Therefore
\begin{eqnarray}
\delta \langle A_t,QA_\eta\rangle_L\lineup = \langle \nabla_t A_\delta,QA_\eta\rangle_L- \langle A_{\delta},\nabla_t \nabla_\eta A_Q\rangle_L,\nonumber\\
\lineup = \langle \nabla_t A_\delta,QA_\eta\rangle_L+ \langle A_{\delta},\nabla_t Q A_\eta\rangle_L,\nonumber\\
\lineup = \frac{d}{dt}\langle A_\delta,QA_\eta\rangle_L.
\end{eqnarray}
Integrating $t$ from $0$ to $1$, we obtain
\begin{equation}\delta S_A = - \left.\langle A_\delta,QA_\eta\rangle_L\right|_{t=1}.\label{eq:WZWvar}\end{equation}
This is the expected variation of a WZW-like action.

We would now like to use this formula to prove that the lifted $A_\infty$ theory has the expected gauge invariances. First consider the $\eta$ gauge invariance \eq{deltaeta}. For this transformation, the potential $A_\delta$ takes the form
\begin{eqnarray}
A_{\delta_\eta}|_{t=1} \lineup = \pi_1\G \frac{1}{1-\eta\PhiA}\otimes \eta\Omega_A\otimes \frac{1}{1-\eta\PhiA},\nonumber\\
\lineup =  \pi_1\G \n \frac{1}{1-\eta\PhiA}\otimes \Omega_A\otimes \frac{1}{1-\eta\PhiA},\nonumber\\
\lineup =  \pi_1(\n - \m_2)\G \frac{1}{1-\eta\PhiA}\otimes \Omega_A\otimes \frac{1}{1-\eta\PhiA},\nonumber\\
\lineup = \left.\nabla_\eta\Omega_B\right|_{t=1},
\end{eqnarray}
where for later reference we define
\begin{equation}\Omega_B\equiv \pi_1\G \frac{1}{1-\eta\PhiA}\otimes\Omega_A\otimes \frac{1}{1-\eta\PhiA}. \label{eq:OmB}\end{equation}
Plugging in we find
\begin{equation}\delta_\eta S_A = - \left.\langle \nabla_\eta \Omega_B ,QA_\eta\rangle_L\right|_{t=1}=0, \end{equation}
since $Q A_\eta$ is annihilated by $\nabla_\eta$. This proves that the action is invariant under this gauge symmetry. Next consider the BRST gauge invariance \eq{deltaQ}. For this transformation, the 
potential $A_\delta$ takes the form
\begin{eqnarray}
A_{\delta_Q}|_{t=1} \lineup = \pi_1\G \frac{1}{1-\eta\PhiA}\otimes \left(\pi_1 \M \frac{1}{1-\eta\PhiA}\otimes\Lambda_A\otimes\frac{1}{1-\eta\PhiA}\right)\otimes \frac{1}{1-\eta\PhiA},\nonumber\\
\lineup =  \left(\pi_1\G \M \frac{1}{1-\eta\PhiA}\otimes\Lambda_A\otimes\frac{1}{1-\eta\PhiA}\right)-\left(\pi_1\G \frac{1}{1-\eta\PhiA}\otimes \left(\pi_1 \M \frac{1}{1-\eta\PhiA}\right)\otimes \frac{1}{1-\eta\PhiA}\otimes\Lambda_A\otimes\frac{1}{1-\eta\PhiA}\right)\nonumber\\
\lineup\ \ \ -\left(\pi_1\G \frac{1}{1-\eta\PhiA}\otimes\Lambda_A\otimes\frac{1}{1-\eta\PhiA}\otimes \left(\pi_1 \M \frac{1}{1-\eta\PhiA}\right)\otimes \frac{1}{1-\eta\PhiA}\right),\nonumber\\
\lineup =  Q\left(\pi_1\G \frac{1}{1-\eta\PhiA}\otimes\Lambda_A\otimes\frac{1}{1-\eta\PhiA}\right)-\left(\pi_1\G \frac{1}{1-\eta\PhiA}\otimes \left(\pi_1 \M \frac{1}{1-\eta\PhiA}\right)\otimes \frac{1}{1-\eta\PhiA}\otimes\Lambda_A\otimes\frac{1}{1-\eta\PhiA}\right)\nonumber\\
\lineup\ \ \ -\left(\pi_1\G \frac{1}{1-\eta\PhiA}\otimes\Lambda_A\otimes\frac{1}{1-\eta\PhiA}\otimes \left(\pi_1 \M \frac{1}{1-\eta\PhiA}\right)\otimes \frac{1}{1-\eta\PhiA}\right),\nonumber\\
\lineup = Q\Lambda_B -\mu_B,\label{eq:AdQ}
\end{eqnarray}
where we have defined
\begin{eqnarray}
\Lambda_B\lineup \equiv \pi_1\G \frac{1}{1-\eta\PhiA}\otimes\Lambda_A\otimes\frac{1}{1-\eta\PhiA},\label{eq:LambdaB}\\
\mu_B\lineup \equiv \pi_1\G \left(\frac{1}{1-\eta\PhiA}\otimes \left(\pi_1 \M \frac{1}{1-\eta\PhiA}\right)\otimes \frac{1}{1-\eta\PhiA}\otimes\Lambda_A\otimes\frac{1}{1-\eta\PhiA}\right.\nonumber\\
\lineup\ \ \ \ \ \ \ \ \ \ \ \left.+ \frac{1}{1-\eta\PhiA}\otimes\Lambda_A\otimes\frac{1}{1-\eta\PhiA}\otimes \left(\pi_1 \M \frac{1}{1-\eta\PhiA}\right)\otimes \frac{1}{1-\eta\PhiA}\right).\label{eq:DeltaT}
\end{eqnarray}
The $Q\Lambda_B$ term vanishes when contracted with $QA_\eta|_{t=1}$ since $Q$ is nilpotent. Substituting the $\mu_B$ term into the variation, while switching to the degree grading, we obtain the expression:
\begin{eqnarray}
\delta_Q S_A \lineup = -\langle \omega_L|\left[\pi_1\G \frac{1}{1-\eta\PhiA}\otimes \left(\pi_1 \M \frac{1}{1-\eta\PhiA}\right)\otimes \frac{1}{1-\eta\PhiA}\otimes\Lambda_A\otimes\frac{1}{1-\eta\PhiA}\right]\nonumber\\
\lineup\ \ \ \ \ \ \ \ \ \ \ \ \ \ \ \ \ \ \otimes\left[\pi_1\G \frac{1}{1-\eta\PhiA}\otimes\left(\pi_1\M \frac{1}{1-\eta\PhiA}\right)\otimes  \frac{1}{1-\eta\PhiA}\right]\nonumber\\
\lineup \ \ \  -\langle \omega_L|\left[\pi_1\G \frac{1}{1-\eta\PhiA}\otimes\left(\pi_1\M \frac{1}{1-\eta\PhiA}\right)\otimes  \frac{1}{1-\eta\PhiA}\right]\nonumber\\
\lineup\ \ \ \ \ \ \ \ \ \ \ \ \ \ \ \ \ \ \otimes\left[\pi_1\G \frac{1}{1-\eta\PhiA}\otimes\Lambda_A \otimes \frac{1}{1-\eta\PhiA}\otimes\left(\pi_1 \M \frac{1}{1-\eta\PhiA}\right)\otimes\frac{1}{1-\eta\PhiA}\right],\label{eq:dQvar}
\end{eqnarray}
where we substituted
\begin{equation}QA_\eta|_{t=1} =\pi_1\G \frac{1}{1-\eta\PhiA}\otimes\left(\pi_1\M \frac{1}{1-\eta\PhiA}\right)\otimes  \frac{1}{1-\eta\PhiA},\end{equation}
and used graded antisymmetry of the symplectic form. Equation \eq{dQvar} can be further simplified to 
\begin{equation}
\delta_Q S_A = -\langle \omega_L|\pi_2 \G \left[\frac{1}{1-\eta\PhiA}\otimes \left(\pi_1 \M \frac{1}{1-\eta\PhiA}\right)\otimes \frac{1}{1-\eta\PhiA}\otimes\Lambda_A\otimes\frac{1}{1-\eta\PhiA}\otimes\left(\pi_1\M \frac{1}{1-\eta\PhiA}\right)\otimes  \frac{1}{1-\eta\PhiA}\right].\ \ \label{eq:Id2}
\end{equation}
We will prove this in appendix \ref{app:cyclic}. Note that we can drop the factor of $\G$ from \eq{Id2} because $\G$ is cyclic. Then the projector $\pi_2$ acts directly on the object in parentheses. However, the object in parentheses has no two-string component---it contains at minimum a tensor product of three string fields. Therefore
\begin{equation}\delta_Q S_A = 0.\end{equation}
and the action contains the expected BRST gauge symmetry.

\subsection{Higher Potentials and the Maurer-Cartan Equation}
\label{subsec:MC}

We would now like to get a better understanding of the nonvanishing field strengths of the lifted $A_\infty$ theory.  We already know that they are covariantly constant in the $\eta$ direction. But since $F_{\eta\eta}$ vanishes, the covariant derivative $\nabla_\eta$ is nilpotent. In fact, it turns out that the field strengths are trivial in the $\nabla_\eta$ cohomology:
\begin{eqnarray}
F_{t\delta}\lineup = -\nabla_\eta A_{t\delta},\nonumber\\
F_{tQ}\lineup = -\nabla_\eta A_{tQ},\nonumber\\
F_{\delta Q}\lineup = -\nabla_\eta A_{\delta Q},\nonumber\\
F_{QQ}\lineup = -\nabla_\eta A_{QQ}.
\end{eqnarray}
More generally,
\begin{equation}F_{ij} = (-1)^{(\eps(i)+1)\eps(j)+1}\nabla_\eta A_{ij},\label{eq:FA}\end{equation}
where the sign is chosen for later convenience. The string field $A_{ij}$ will be called the {\it 2-potential}, and is given by the formula
\begin{eqnarray}
A_{ij}(t) \lineup \equiv \pi_1\G\left[(-1)^{(\deg(i)+1)(\deg(j) +1)}\frac{1}{1-\eta\PhiA(t)}\otimes a_i(t)\otimes \frac{1}{1-\eta\PhiA(t)}\otimes a_j(t)\otimes \frac{1}{1-\eta\PhiA(t)}\right.\nonumber\\
\lineup\ \ \ \ \ \ \ \ \ \ \ \ \ \ \ \ \ \ \ \ \ \ \ \ \ \ \ \ \ \ \ \ \ \ \ \ \ \
+ \frac{1}{1-\eta\PhiA(t)}\otimes a_j(t)\otimes \frac{1}{1-\eta\PhiA(t)}\otimes a_i(t)\otimes \frac{1}{1-\eta\PhiA(t)}\nonumber\\
\lineup \left.\ \ \ \ \ \ \ \ \ \ \ \ \ \ \ \ \ \ \ \ \ \ \ \ \ \ \ \ \ \ \ \ \ \ \ \ \ \
+\frac{1}{1-\eta\PhiA(t)}\otimes a_{ij}(t)\otimes \frac{1}{1-\eta\PhiA(t)}\right].\ \ \ \label{eq:Aij}
\end{eqnarray}
The string field $a_{ij}(t)$ will be called the {\it little 2-potential}, and is defined
\begin{eqnarray}
a_{ij}(t)\lineup \equiv \int_0^t ds\Bigg[ (-1)^{(\deg(i)+1)(\deg(j)+1)}\pi_1\D_i\left(\frac{1}{1\!-\!\eta\PhiA(s)}\otimes\dot{\Phi}_A(s)\otimes\frac{1}{1\!-\!\eta\PhiA(s)}\otimes a_{j}(s)\otimes \frac{1}{1\!-\!\eta\PhiA(s)}\right.\nonumber\\
\lineup \ \ \ \ \ \ \ \ \ \ \ \ \ \ \ \ \ \ \ \ \ \ \ \ \ \ \ \ \ \ \ \ \ \left.+(-1)^{\deg(j)+1} \frac{1}{1\!-\!\eta\PhiA(s)}\otimes a_j(s)\otimes \frac{1}{1\!-\!\eta\PhiA(s)}\otimes \dot{\Phi}_A(s)\otimes \frac{1}{1\!-\!\eta\PhiA(s)}\right)\nonumber\\
\lineup\ \ \  \ \ \ \ \ \ \ \ \ \ \ \ \ \ \ \ \ \ \ \ \ \ \ \ \ \ \ \ \ \ \ \ \ \ \ \ + \pi_1\D_{j}\left(\frac{1}{1\!-\!\eta\PhiA(s)}\otimes\dot{\Phi}_A(s)\otimes\frac{1}{1\!-\!\eta\PhiA(s)}\otimes a_i(s)\otimes \frac{1}{1\!-\!\eta\PhiA(s)}\right.\nonumber\\
\lineup \ \ \ \ \ \ \ \ \ \ \ \ \ \ \ \ \ \ \ \ \ \ \ \ \ \ \ \ \ \ \ \ \ \left.+(-1)^{\deg(i)+1} \frac{1}{1\!-\!\eta\PhiA(s)}\otimes a_i(s)\otimes \frac{1}{1\!-\!\eta\PhiA(s)}\otimes \dot{\Phi}_A(s)\otimes \frac{1}{1\!-\!\eta\PhiA(s)}\right)\Bigg].\ \ \ \ \ \ \ 
\end{eqnarray}
Interestingly, the 2-potential $A_{ij}$ looks like a two-index generalization of the potential $A_i$ given in equation \eq{Ad}, and the little 2-potential $a_{ij}$ looks like a two-index generalization of the little potential $a_i$ given in equation \eq{ad}. Computing $\nabla_\eta A_{ij}$ reproduces the field strength $F_{ij}$ as a result of the identity
\begin{eqnarray}
0= \eta a_{ij}(t) \lineup +  (-1)^{(\deg(i)+1)\deg(j)}\pi_1\D_i \frac{1}{1\!-\!\eta\PhiA(t)}\otimes a_j(t)\otimes \frac{1}{1\!-\!\eta\PhiA(s)}\nonumber\\
\lineup \ \ \ \ \ \ \ \ \,  +(-1)^{\deg(j)+1}\pi_1\D_j\frac{1}{1\!-\!\eta\PhiA(t)}\otimes a_i(t)\otimes \frac{1}{1\!-\!\eta\PhiA(t)}, \label{eq:little2}
\end{eqnarray}
which is a kind of two-index generalization of the identity \eq{little1}. The 2-potentials have not played a role so far since they do not appear in the action. But there are other expressions for the action where 2-potentials do appear. For example, if we try to express the action in the standard WZW-like form \eq{stWZW}, we find an additional term proportional to the 2-potential $A_{tQ}$: 
\begin{equation}
S_A = -\frac{1}{2}\langle A_\eta,A_Q\rangle_L\Big|_{t=1} - \frac{1}{2}\int_0^1 dt\, \langle A_t,[A_\eta,A_Q]\rangle_L - \frac{1}{2}\int_0^1 dt \langle A_{tQ}, A_\eta*A_\eta\rangle_L.
\end{equation}
A similar generalization of the standard WZW-like action has also been discussed in the context of heterotic string field theory \cite{BOZ}. 

It turns out that the story does not end with 2-potentials. By factoring $\nabla_\eta$ out of the Bianchi identity for the nonvanishing field strengths, we learn that the 2-potentials must satisfy the identity:
\begin{eqnarray}
\lineup \nabla_i A_{jk} + (-1)^{\eps(j)(\eps(i)+\eps(k)) +\eps(j)+\eps(k)}\nabla_j A_{ki} +(-1)^{\eps(i)(\eps(j)+\eps(k)) +\eps(i)+\eps(j)}\nabla_kA_{ij}\nonumber\\
\lineup\ \ \ \ \ \ \ \ \ \ \ \ \ \ \ \ \ \ \ \ \ \ \ \ \ \ \ \ \ \ \ \ \ \ \ \ \ \ \ \ \ \ \ \ \ \ \ \ \ \ \ \ \ \ \ \ \ \ \ \ \ \ \ \ \ \ \ \ \ \ \ \ \ \ \ \ \ \ \ \ \ \ 
+(-1)^{(\eps(i)+1)(\eps(j)+\eps(k)+1)}\nabla_\eta A_{ijk}=0, \ \ \ \ \ \ \ \label{eq:Bianchi1}
\end{eqnarray}
where $A_{ijk}$ is a new object called the {\it 3-potential}. Acting on this equation with the covariant derivative $\nabla_i$ and symmetrizing, one can further introduce a {\it 4-potential}, and so on. To clarify the structure of the hierarchy, it is helpful to invoke the language of differential forms.\footnote{The author would like to thank S. Konopka for discussion which clarified the nature of this hierarchy.} For each derivation $\d_i$ we formally introduce a corresponding basis 1-form:
\begin{equation}\d_i\leftrightarrow dx^i.\end{equation}
Note that we do not introduce a 1-form dual to $\eta$. The basis 1-forms carry Grassmann parity and degree
\begin{equation}
\eps(dx^i) \equiv \deg(dx^i) \equiv \eps(i)+1,\ \ \ \mathrm{mod}\ \mathbb{Z}_2,
\end{equation}
and ghost and picture number
\begin{eqnarray}
\mathrm{gh}(dx^i)\lineup  \equiv 1-\mathrm{gh}(\d_i),\nonumber\\
\mathrm{picture}(dx^i) \lineup \equiv -1-\mathrm{picture}(\d_i).
\end{eqnarray}
We consider the algebra of ``string field-valued" differential forms, defined in the obvious way by tensoring the wedge product with the open string star product, with the appropriate signs from (anti)commutation of string fields and forms. We define the exterior derivative
\begin{equation}d \equiv dx^i \d_i.\end{equation}
The exterior derivative is nilpotent and commutes with $\eta$, since the $\d_i$s commute among themselves and with $\eta$. Also, $d$ acts as a derivation of the wedge/star product. Next, we introduce differential forms corresponding to the $n$-potentials:\footnote{The field strength as defined in \eq{F} does not have the right symmetry properties in the odd directions to define a 2-form. This can be fixed by multiplying the field strength by the appropriate sign, but we did not bother since it would contradict the definition used in previous papers. This is the origin of the sign factor relating the 2-potential and the field strength in \eq{FA}.}
\begin{equation}
A^{(0)} \equiv A_\eta,\ \ \ \ \ \ A^{(1)} \equiv dx^i A_i,\ \ \ \ \ A^{(2)} \equiv \frac{1}{2!}dx^i\wedge dx^j A_{ij},\ \ \ \ A^{(3)} \equiv \frac{1}{3!}dx^i \wedge dx^j\wedge dx^k A_{ijk},\ \ ...\ .
\end{equation}
Note that $A_\eta$ is a scalar, while $A_i$ are components of a 1-form. Therefore $A_\eta$ can be interpreted as the {\it 0-potential} and $A_i$ as the {\it 1-potential}, and the pattern completes to higher potentials. Accounting for the Grassmannality, ghost and picture number of the basis 1-forms, the $n$-potentials and exterior derivative are Grassmann odd, carry ghost number 1 and picture $-1$. On the basis of previous calculations it is easy to check that the potentials satisfy
\begin{eqnarray}
0\lineup = \eta A^{(0)} - A^{(0)}*A^{(0)},\\
0\lineup = d A^{(0)} +\eta A^{(1)} -[A^{(0)},A^{(1)}],\\
0\lineup = dA^{(1)} +\eta A^{(2)} - [A^{(0)},A^{(2)}] -A^{(1)}*A^{(1)},\\
0\lineup = dA^{(2)} +\eta A^{(3)}- [A^{(0)},A^{(3)}] -[A^{(1)},A^{(2)}],\\
\lineup \vdots\ \ .\nonumber
\end{eqnarray}
The first two identities are equivalent to the statement that the field strengths vanish in the $\eta$ direction. The third is equivalent to the statement that the nonvanishing field strengths are trivial in the $\nabla_\eta$ cohomology, and the fourth is implied by factoring $\nabla_\eta$ out of the Bianchi identity. It is clear that these identities arise as components of a Maurer-Cartan equation:
\begin{equation}(d + \eta) A - A*A = 0,\end{equation}
where $A$ is given by the formal sum 
\begin{equation}A \equiv A^{(0)}+A^{(1)}+A^{(2)}+A^{(3)}+...\ .\end{equation}
We will call $A$ the {\it multi-potential}. The Maurer-Cartan equation can be thought of as an ``equation of motion" which determines the potentials and their higher-form descendants as functionals of the interpolation $\Phi(t)$. Once we have solved this equation, we can write down a WZW-like action. Actually, for this purpose we need to assume an additional regularity condition: An arbitrary variation of the dynamical field $\Phi$ produces an arbitrary $A_\delta$ at $t=1$:
\begin{equation}
\mathrm{Regularity\ Condition}:\ \ \ \mathrm{arbitrary}\ \delta\Phi\ \to\ \mathrm{arbitrary}\ A_\delta|_{t=1}.
\end{equation}
This needs to be true, otherwise the variation of the WZW-like action \eq{WZWvar} does not imply the correct equations of motion.  Thus, for example, $A=0$ would not be a regular solution for the purposes of defining a WZW-like action. At least perturbatively, the potential $B_\delta$ of the Berkovits theory and the potential $A_\delta$ of the lifted $A_\infty$ theory are regular, since, to leading order in the string field, they are proportional to $\delta\PhiB(t)$ and $\delta \PhiA(t)$, respectively.

Berkovits' superstring field theory provides a solution to the Maurer-Cartan equation in the form
\begin{equation}
B = B_\eta + dx^i B_i.
\end{equation}
Since the field strengths vanish, the higher potentials can be set to zero. In the lifted $A_\infty$ theory the solution to the Maurer-Cartan equation is more interesting. We introduce differential forms corresponding to the little potentials:
\begin{equation}
a^{(0)} \equiv \eta\PhiA(t),\ \ \ \ \ a^{(1)} \equiv dx^i a_i,\ \ \ \  a^{(2)} \equiv \frac{1}{2!}dx^i\wedge dx^j a_{ij},\ \ \  a^{(3)} \equiv \frac{1}{3!}dx^i \wedge dx^j\wedge dx^k a_{ijk},\ \  ... \ .
\end{equation}
For convenience we have defined the {\it little 0-potential} to be $a^{(0)}=\eta\PhiA(t)$. All little $n$-potentials are degree even, ghost number 1 and picture $-1$ once we account for the basis 1-forms. We define the {\it little multi-potential} as the formal sum
\begin{equation}
a \equiv a^{(0)}+a^{(1)}+a^{(2)}+a^{(3)}+...,
\end{equation}
and postulate a solution to the Maurer-Cartan equation in the form 
\begin{equation}A = \pi_1\G \frac{1}{1-a}.\label{eq:Aa}\end{equation}
Here the tensor algebra of ``string-field-valued" differential forms is defined in the obvious way by tensoring the wedge product of the basis 1-forms with the tensor product of string fields, with the appropriate signs from (anti)commutation of the string fields and forms. Note that \eq{Aa} agrees with our earlier expressions for $A_\eta,A_i,A_{ij}$ once we extract the zero, one and two form components. The Maurer-Cartan equation for the multi-potential $A$ translates into an equation for the little multi-potential $a$: 
\begin{equation}
\pi_1(\n+\D)\frac{1}{1-a} = 0, \label{eq:littleMC}
\end{equation}
where $\D \equiv dx^i \D_i$. Note that this produces \eq{little1} and \eq{little2} when we extract the 1-form and 2-form components. The solution to this equation is not unique, but extrapolating from the expressions for $a_i$ and $a_{ij}$, we propose a solution of the form
\begin{equation}
a(t) = \int_0^t ds\, \pi_1(\n+ \D)\frac{1}{1-a(s)}\otimes\dot{\Phi}_A(s)\otimes\frac{1}{1-a(s)}.\label{eq:littlesol}
\end{equation}
This defines the little potentials and their higher-form descendants recursively; the $n$-form component determines the little $n$-potential in terms of products of little $k$-potentials for $k<n$. To show that this formula solves \eq{littleMC}, take the $t$ derivative of the left hand side of \eq{littleMC} and substitute
\begin{equation}
\dot{a}(t) = \pi_1(\n+ \D)\frac{1}{1-a(t)}\otimes\dot{\Phi}_A(t)\otimes\frac{1}{1-a(t)}.
\end{equation}
This gives
\begin{eqnarray}
\frac{d}{dt}\pi_1(\n+\D)\frac{1}{1-a}\lineup = \pi_1(\n+\D)\frac{1}{1-a}\otimes\left( \pi_1(\n+ \D)\frac{1}{1-a}\otimes\dot{\Phi}_A(t)\otimes\frac{1}{1-a}\right)\otimes\frac{1}{1-a},\nonumber\\
\lineup = \pi_1(\n+\D)^2 \frac{1}{1-a}\otimes\dot{\Phi}_A(t)\otimes\frac{1}{1-a},\nonumber\\
\lineup \ \ \ -\pi_1(\n+\D)\frac{1}{1-a}\otimes\left(\pi_1(\n+\D)\frac{1}{1-a}\right)\otimes\frac{1}{1-a}\otimes\dot{\Phi}_A(t)\otimes\frac{1}{1-a}\nonumber\\
\lineup\ \ \ +\pi_1(\n+\D)\frac{1}{1-a}\otimes \dot{\Phi}_A(t)\otimes\frac{1}{1-a}\otimes\left(\pi_1(\n+\D)\frac{1}{1-a}\right)\otimes\frac{1}{1-a}.
\end{eqnarray}
Note that $(\D+\n)^2$ vanishes since $d$ and $\eta$ are nilpotent and mutually commuting derivations of the star product. Bringing the last two terms to the other side of the equation gives
\begin{eqnarray}
0=\frac{d}{dt}\pi_1(\n+\D)\frac{1}{1-a} \lineup+ \pi_1(\n+\D)\frac{1}{1-a}\otimes\left(\pi_1(\n+\D)\frac{1}{1-a}\right)\otimes\frac{1}{1-a}\otimes\dot{\Phi}_A\otimes\frac{1}{1-a}\nonumber\\
\lineup-\pi_1(\n+\D)\frac{1}{1-a}\otimes \dot{\Phi}_A\otimes\frac{1}{1-a}\otimes\left(\pi_1(\n+\D)\frac{1}{1-a}\right)\otimes\frac{1}{1-a}.
\end{eqnarray}
This is a first order homogeneous differential equation in the string field $\pi_1 (\n+\D)\frac{1}{1-a}$. Since $\pi_1 (\n+\D)\frac{1}{1-a}$ vanishes at $t=0$ (because $\PhiA(t)$ vanishes at $t=0$), it must vanish for all $t$. Therefore \eq{littleMC} is satisfied, and \eq{Aa} gives a solution to the Maurer-Cartan equation. 

One advantage of the Maurer-Cartan equation is that it gives a clearer understanding of the symmetries implicit in the choice of potentials used to express the action in WZW-like form. In particular, solutions can be modified by an infinitesimal ``gauge transformation"
\begin{equation}
\delta_{mc} A = (d+\eta)\Lambda -[A,\Lambda],
\end{equation}
where $\Lambda$ is a sum of $n$-form gauge parameters
\begin{equation}\Lambda = \Lambda^{(0)} + dx^i \Lambda_i + \frac{1}{2!}dx^i\wedge dx^j \Lambda_{ij}+ \frac{1}{3!}dx^i \wedge dx^j\wedge dx^k \Lambda_{ijk}+...\ . \end{equation}
In general $\Lambda$ can be a functional of the interpolation $\Phi(t)$, and is Grassmann even, ghost and picture number zero. Note that this ``gauge transformation" alters the choice of the potentials---not the dynamical string field on which the potentials implicitly depend. It is interesting to see how this transformation effects the WZW-like action. For this purpose it is enough to consider how the 0- and 1-potentials transform:
\begin{eqnarray}
\delta_{mc} A_\eta\lineup  = \nabla_\eta \Lambda^{(0)},\nonumber\\
\delta_{mc} A_i \lineup = \nabla_i \Lambda^{(0)} + \nabla_\eta \Lambda_i.
\end{eqnarray}
Following a similar computation as in the previous section, it is not difficult to show that the WZW-like action changes as
\begin{equation}
\delta_{mc}\int_0^1 dt\langle A_t,QA_\eta\rangle_L = \langle \Lambda^{(0)}, Q A_\eta\rangle_L|_{t=1}.
\end{equation}
From this we make two observations. First, while the 1-form parameter $\Lambda_i$ alters the potentials, it does not change the action. Second, the 0-form parameter $\Lambda^{(0)}$ changes the action in the same way as a variation of the field.  Thus the action after the transformation is the same as the original action with the replacement $\Phi\to\Phi+\delta\Phi$, where $\delta\Phi$ is determined by solving the equation 
\begin{equation}A_\delta = \Lambda^{(0)}.\end{equation}
Thus the Maurer-Cartan gauge transformation alters the WZW-like action at most by a field redefinition, and then only if the zero-form component of the transformation parameter is nonvanishing. Furthermore, it is clear that a Maurer-Cartan gauge transformation of the form
\begin{equation}\Lambda^{(0)}|_{t=1} = Q\Lambda_B + \nabla_\eta \Omega_B \end{equation}
will leave the action invariant. By equating this with $A_\delta$, this determines an infinitesimal gauge transformation of the fields. In this sense, the Mauer-Cartan equation plays a role for the WZW-like action somewhat analogous to the role of cyclic-$A_\infty$ algebras in the $A_\infty$ action: It is an algebraic structure that is covariant under field redefinition and implies the existence of a gauge invariant action.

The interpretation of the WZW-like action suggested by the Maurer-Cartan equation is slightly different from the earlier interpretation in terms of potentials and field strengths. For example, from the Maurer-Cartan perspective $A_\eta$ is a scalar, and so should not define components of a field strength. Nevertheless, the statement that the field strengths vanish in the $\eta$ direction is a convenient way to characterize the essential information in the Maurer-Cartan equation for the purposes of writing the action. We will translate back and forth between both interpretations as convenient.

\subsection{Little Potentials}
\label{subsec:little}

The action of the lifted $A_\infty$ theory can be formulated in a different way using little potentials. In \eq{action4} the lifted $A_\infty$ action was expressed in the form
\begin{equation}S_{A} = \int_0^1dt\,\omega_L\left(\dot{\Phi}_A(t),\pi_1\M \frac{1}{1-\eta\PhiA(t)}\right).\end{equation}
Note that the little potentials satisfy
\begin{equation}a_t(t) = \dot{\Phi}_A(t),\ \ \ \eta a_Q(t) = -\pi_1\M\frac{1}{1-\eta\PhiA(t)},\end{equation}
where the second equality follows from \eq{little1}. Thus the $A_\infty$ action can be expressed 
\begin{equation}S_A = -\int_0^1 dt\,\omega_L(a_t,\eta a_Q).\end{equation}
This is reminiscent of a WZW-like action, but note that the little potentials do not come from a solution to the Maurer-Cartan equation. Instead, the relevant equation for the little potentials is \eq{littleMC}.

To make the description in terms of little potentials more familiar, we can define a few objects by analogy with the potentials and field strengths of the WZW-like action. First, define the little potential in the $\eta$ direction to be the little 0-potential:
\begin{equation}
a_\eta \equiv a^{(0)} = \eta\Phi_A(t).
\end{equation}
Second, we define a ``little field strength:"
\begin{equation}
f_{ij} \equiv (-1)^{(\deg(i)+1)\deg(j)+1}\eta a_{ij}.
\end{equation}
Unlike the WZW-like formulation, it does not make sense to define components of the little field strength in the $\eta$ direction, for reasons that will be clear in a moment. The little field strength is graded antisymmetric,
\begin{equation}f_{ij} = -(-1)^{\deg(i)\deg(j)} f_{ji},\end{equation}
and constant in the $\eta$ direction:
\begin{equation}\eta f_{ij}=0.\end{equation}
Recall that the potentials and field strengths satisfy  \eq{dswitch} 
\begin{equation}\d_I A_{J} = (-1)^{\eps(I)\eps(J)}\nabla_{J} A_{I}+F_{IJ}.\label{eq:dswitch2}\end{equation}
Little potentials satisfy analogous identities which allow us to swap the index of a coderivation with the index of the little potential. In particular, \eq{little2} can be expressed
\begin{equation}
\pi_1\D_i\frac{1}{1-a_\eta}\otimes a_j\otimes \frac{1}{1-a_\eta}
=(-1)^{\deg(i)\deg(j)} \pi_1\D_j\frac{1}{1-a_\eta}\otimes a_i\otimes \frac{1}{1-a_\eta}\, +\, f_{ij},\label{eq:littleswitch1}
\end{equation}
where switching $i$ and $j$ adds the little field strength $f_{ij}$. Meanwhile \eq{little1} can be expressed 
\begin{equation}
\eta a_i  = (-1)^{\deg(i)} \pi_1\D_i\frac{1}{1-a_\eta}.\label{eq:littleswitch2}
\end{equation}
Note that this last identity is not a special case of the previous one after choosing $\D_i$ to be $\eta$ and setting $f_{ij}$ to zero. This is why it doesn't make sense to define components of the little field strength in the $\eta$ direction. Except for this small difference, the formulation is quite similar to that of the WZW-like action.

Now let's compute the variation of the action. Take the variation of the integrand:
\begin{eqnarray}
\delta\, \omega_L(a_t,\eta a_Q)\lineup =\omega_L(\delta a_t,\eta a_Q) + \omega_L(a_t,\eta\delta a_Q),\nonumber\\
\lineup = \omega_L(\dot{a}_\delta + f_{\delta t},\eta a_Q) + \omega_L\Bigg(a_t,\eta\left(\pi_1\M\frac{1}{1-a_\eta}\otimes a_\delta \otimes\frac{1}{1-a_\eta} +f_{\delta Q}\right)\Bigg),
\end{eqnarray}
where in the second step we used \eq{littleswitch1}. Note that the identities \eq{littleswitch1} and \eq{littleswitch2} simplify quite a bit in the case of the derivations $\delta$ and $d/dt$ since they commute with the cohomomorphism $\G$. We can drop the little field strengths $f_{\delta t}$ and $f_{\delta Q}$ since they are constant in the $\eta$ direction (in fact $f_{\delta t}$ already vanishes identically). Moving $\eta$ in the second term onto the first entry of the symplectic form and using \eq{littleswitch2} gives
\begin{eqnarray}
\delta\, \omega_L(a_t,\eta a_Q)\lineup = \omega_L(\dot{a}_\delta,\eta a_Q) + \omega_L\left(\dot{a}_\eta, \pi_1\M\frac{1}{1-a_\eta}\otimes a_\delta \otimes\frac{1}{1-a_\eta}\right).
\end{eqnarray}
Using cyclicity of $\M$ one can show that 
\begin{equation}
\omega_L\left(\dot{a}_\eta, \pi_1\M\frac{1}{1-a_\eta}\otimes a_\delta \otimes\frac{1}{1-a_\eta}\right) = -\omega_L\left(\pi_1\M\frac{1}{1-a_\eta}\otimes \dot{a}_\eta \otimes\frac{1}{1-a_\eta}, a_\delta\right). \label{eq:Id3}
\end{equation}
We prove this in appendix \ref{app:cyclic}. Therefore 
\begin{eqnarray}
\delta\, \omega_L(a_t,\eta a_Q)\lineup = \omega_L(\dot{a}_\delta,\eta a_Q) - \omega_L\left(\pi_1\M\frac{1}{1-a_\eta}\otimes \dot{a}_\eta \otimes\frac{1}{1-a_\eta}, a_\delta\right),\nonumber\\
\lineup = \omega_L(\dot{a}_\delta,\eta a_Q) - \omega_L\left(\frac{d}{dt}\pi_1\M\frac{1}{1-a_\eta}, a_\delta\right),\nonumber\\
\lineup =  \omega_L(\dot{a}_\delta,\eta a_Q) + \omega_L\left(\eta \dot{a}_Q , a_\delta\right),\nonumber\\
\lineup = \frac{d}{dt}\omega_L(a_\delta,\eta a_Q).
\end{eqnarray}
Integrating from $0$ to $1$ gives the variation of the action
\begin{equation}
\delta S_A = -\omega_L(a_\delta,\eta a_Q)|_{t=1}.
\end{equation}
This implies that the equations of motion can be expressed 
\begin{equation}\eta a_Q|_{t=1} = 0,\end{equation}
which using \eq{littleswitch2} is equivalent to 
\begin{equation}\pi_1\M\frac{1}{1-\PsiA} = 0.\end{equation}
This is the standard expression for the $A_\infty$ equations of motion. By contrast the WZW-like action gives the equations of motion in a ``Berkovits-like" form  
\begin{equation} 
Q \left(\pi_1\G\frac{1}{1-\PsiA}\right) = 0.
\end{equation}
These two formulations are related by conjugation with the cohomomorphism $\G$.

\section{Field Redefinition}

We now search for a field redefinition relating the dynamical field $\PhiA$ of the lifted $A_\infty$ theory to the dynamical field $\PhiB$ of the Berkovits theory. This field redefinition should transform the WZW-like action of the lifted $A_\infty$ theory
\begin{equation}S_A = - \int_0^1dt\, \langle A_t,QA_\eta\rangle_L,\end{equation}
into to the WZW-like action of the Berkovits theory
\begin{equation}S_B = - \int_0^1dt\, \langle B_t,QB_\eta\rangle_L,\end{equation}
where the potentials $B_t$ and $B_\eta$ for the Berkovits theory were defined in \eq{Bpotentials}.

In earlier work, the relation between the Berkovits and $A_\infty$ theories was described by partially gauge fixing the Berkovits theory to the small Hilbert space. In this context, the field redefinition between the theories was given by equating the respective $\eta$ potentials \cite{OkWB}:
\begin{equation}B_\eta|_{t=1}=A_\eta|_{t=1}.\label{eq:AnBn}\end{equation}
Unfortunately this condition is not enough to specify a field redefinition between $\PhiA$ and $\PhiB$ in the large Hilbert space. It leaves is a residual ambiguity related to an overall $\eta$ gauge transformation. There are a number of ways to fix this ambiguity. We will take a particular approach which at first seems orthogonal but has interesting implications. Instead of equating the $\eta$ potentials, we will equate the $t$ potentials:
\begin{equation}A_t = B_t.\label{eq:AtBt}\end{equation}
In doing this, we cannot only be talking about the fields $\PhiA$ and $\PhiB$ at $t=1$. In fact, in a sense we will describe, this equation provides a field redefinition between the entire interpolations $\PhiA(t)$ and $\PhiB(t)$. 

\subsection{From $A_\infty$ to Berkovits}

Let us solve equation \eq{AtBt}. Substituting $B_t$ we obtain a differential equation for $\PhiB(t)$:
\begin{equation}\frac{d}{dt} e^{\PhiB(t)} =  A_t\, e^{\PhiB(t)}.\label{eq:AtBtdiff}\end{equation}
The solution is provided by a {\it Wilson line}
\begin{equation}g(t_2,t_1)=\overleftarrow{\mathcal{P}} \exp\left[\int_{t_1}^{t_2} ds\, A_t(s)\right],\end{equation}
where $\overleftarrow{\mathcal{P}}$ denotes the path ordered exponential in sequence of {\it decreasing} $s$. To emphasize this, we have written $g(t_2,t_1)$ so that the left-most argument $t_2$ is viewed as a {\it later} time than the right-most argument $t_1$. Thus we have 
\begin{equation}e^{\PhiB(t)}=g(t,0) = \overleftarrow{\mathcal{P}}\exp\left[\int_{0}^{t} ds\, A_t(s)\right],\label{eq:BAt}\end{equation}
and, at $t=1$ 
\begin{equation}e^{\PhiB}=g(1,0) = \overleftarrow{\mathcal{P}} \exp\left[\int_{0}^{1} ds\, A_t(s)\right].\label{eq:BA}\end{equation}
This is our proposed field redefinition between $\PhiA$ and $\PhiB$.

To make sense of this field redefinition we must clarify an important point of interpretation. At face value, the Wilson line \eq{BA} expresses $\PhiB$ as a function of the entire path $\PhiA(t)$, not only of $\PhiA$ at $t=1$. Therefore we need to be more specific about the meaning of $\PhiA(t)$ when $t\neq 1$. We will assume $\PhiA(t)$ is given as some time-dependent function of the dynamical field $\PhiA$:
\begin{equation}\PhiA(t) = f_A(t,\PhiA),\end{equation}
which is subject to boundary conditions  
\begin{equation}f_A(0,\PhiA) = 0,\ \ \ f_A(1,\PhiA) = \PhiA.\label{eq:boundaryf}\end{equation}
Under this assumption, then indeed the Wilson line \eq{BA} expresses $\PhiB$ as a function of $\PhiA$. However, for this to be a valid field redefinition, it must be invertible. To simplify analysis of this question, we will make an additional assumption about the interpolating function. The interpolating function will have an expansion in powers of the string field:
\begin{equation}f_A(t,\PhiA) = f_{0}(t) + f_{1}(t)\PhiA + f_{2}(t)\PhiA\otimes\PhiA + ... ,\label{eq:fA}\end{equation}
where the linear maps $f_{n}(t):\mathcal{H}^{\otimes n}\to\mathcal{H}$ are $n$-string products. We will assume that the zero string product in this expansion vanishes:
\begin{equation}f_{0}(t) = 0.\label{eq:fass}\end{equation}
Using this, we can compute the proposed field redefinition out to first order in the fields. To first order, the $t$ potential of the lifted $A_\infty$ theory takes the form
\begin{equation}A_t = \dot{f}_{1}(t)\PhiA + \mathrm{higher\ orders}.\end{equation}
Therefore, to first order, the proposed field redefinition takes the form
\begin{eqnarray}1+\PhiB +\mathrm{higher\ orders} \lineup = 1+ \int_0^1 ds\, \dot{f}_{1}(s)\PhiA + \mathrm{higher\ orders},\nonumber\\
\lineup = 1+ \big(f_{1}(1)-f_{1}(0)\big)\PhiA + \mathrm{higher\ orders}.
\end{eqnarray}
The boundary condition \eq{boundaryf} implies $f_{1}(1) = \mathbb{I}$ and $f_{1}(0) = 0$. Therefore the proposed field redefinition is simply
\begin{equation}\PhiB = \PhiA +\mathrm{higher\ orders}.\end{equation}
This is obviously invertible at first order, and small corrections in higher powers of the field will not change this fact. Therefore, at least perturbatively, the Wilson line \eq{BA} is a valid field redefinition between $\PhiA$ and $\PhiB$. If we relax the assumption \eq{fass}, the proposed field redefinition will map the vacuum configuration $\PhiA = 0$ to a finite, pure gauge configuration for $\PhiB$. The question of invertibility in this case is more complicated, and we will not consider it. 

Now let's return to equation \eq{BAt}, which determines the Berkovits field $\PhiB(t)$ when $t\neq 1$. With a given choice of interpolating function for $\PhiA(t)$, equation \eq{BAt}  expresses $\PhiB(t)$ as a function of $\PhiA$. In turn, we can express $\PhiA$ as a function $\PhiB$ by inverting the field redefinition, so in fact \eq{BAt} gives an interpolating function for the Berkovits theory, 
\begin{equation}\Phi_B(t) = f_B(t,\PhiB),\label{eq:fB}\end{equation}
which implicitly depends on the choice of interpolating function $f_A(t,\PhiA)$ of the lifted $A_\infty$ theory. This is the sense that the equation $A_t=B_t$ provides a ``field redefinition" between the interpolations $\PhiA(t)$ and $\PhiB(t)$.

For the sake of being concrete, let us compute the field redefinition between $\PhiA$ and $\PhiB$ out to second order in the string field. Expanding $e^{\PhiB}$ and the path ordered exponential to second order gives the expression
\begin{eqnarray}
\PhiB \lineup = \PhiA + \int_0^1 ds\, \Big(\mu_2\big(\eta f_1(s)\PhiA,\dot{f}_1(s)\PhiA\big) + \mu_2\big(\dot{f}_1(s)\PhiA,\eta f_1(s)\PhiA\big) +\big(\dot{f}_1(s)\PhiA\big)f_1(s)\PhiA\Big)-\frac{1}{2}\PhiA^2\nonumber\\
\lineup\ \ \ \ \ \ \ \ \ \ \ \ \ \ \ \ \ \ \ \  +\mathrm{higher\ orders},\label{eq:2ndgen}
\end{eqnarray}
where $\mu_2$ is the gauge 2-product of the $A_\infty$ theory \cite{WittenSS} and $f_1(t)$ is the 1-string product in the interpolating function \eq{fA}. Let us restrict to a particular class of field redefinitions which can be written exclusively in terms of $\xi,\eta$ and the open string star product. This implies that $f_1(t)$ can take the form
\begin{equation}f_1(t) = x(t) \mathbb{I} + y(t)\xi\eta,\end{equation}
where $x(t),y(t)$ are number-valued functions of $t$ satisfying boundary conditions $x(0)=y(0)=0$ and $x(1)=1$ and $y(1)=0$. While there are an infinite number of possible choices of $x(t)$ and $y(t)$, reparameterization invariance of the field redefinition implies that most choices are equivalent.  By inspection of \eq{2ndgen}, it is clear that the only reparameterization invariant quantity that can appear at second order is 
\begin{equation}C = \int_0^1 ds\, x(s)\dot{y}(s),\end{equation}
since the other possible combinations $x\dot{x}$ and $y\dot{y}$ are total derivatives which are fixed by boundary conditions. Explicitly, we find that the field redefinition takes the form 
\begin{eqnarray}
\PhiB\lineup = \PhiA+\frac{1}{3}\left(\frac{1}{2}+C\right)\Big(\xi[\eta\PhiA,\PhiA]-[\xi\PhiA,\eta\PhiA]\Big)-\frac{C}{3}\xi[\eta\PhiA,\xi\eta\PhiA]\nonumber\\
\lineup\ \ \ \ \ \ \ \ +\frac{1}{3}\left(\frac{1}{2}-2C\right)[\xi\eta\PhiA,\PhiA]+\mathrm{higher\ orders}.\label{eq:BA2nd}
\end{eqnarray}
If we change the constant $C$, the field redefinition will change by a term proportional to  
\begin{equation}
\xi[\eta\PhiA,\PhiA]-[\xi\PhiA,\eta\PhiA]-\xi[\eta\PhiA,\xi\eta\PhiA]-2[\xi\eta\PhiA,\PhiA].\label{eq:BAgauge}
\end{equation}
One can check that this term is annihilated by $\eta$, and therefore represents an $\eta$ gauge transformation. It is interesting to note that the field redefinition is not completely arbitrary, despite having a free parameter $C$. The most general field redefinition at second order constructed out of $\eta,\xi$, and the star product actually has five free parameters (only four if we require that the fields are equal at linear order). 

Perhaps the most significant appeal of this approach is that it gives the simplest possible proof of equivalence of the actions. All we have to do is prove that the $\eta$ potentials are equal:
\begin{eqnarray}
B_\eta \lineup = (\eta e^{\PhiB(t)})e^{-\PhiB(t)},\nonumber\\
\lineup = \big(\eta g(t,0)\big)g(t,0)^{-1},\nonumber\\
\lineup = \left[\int_0^t ds\, g(t,s)\Big(\eta A_t(s)\Big) g(s,0)\right] g(t,0)^{-1},\nonumber\\
\lineup = \left[\int_0^t ds\, g(t,s)\left(\frac{d}{ds} A_\eta(s)-[A_\eta(s),A_t(s)]\right) g(s,0)\right] g(t,0)^{-1},\nonumber\\
\lineup = \left[\int_0^t ds\, \frac{d}{ds}\Big(g(t,s) A_\eta(s) g(s,0)\Big)\right] g(t,0)^{-1},\nonumber\\
\lineup = A_\eta(t) g(t,0)g(t,0)^{-1},\nonumber\\
\lineup = A_\eta.\label{eq:BnAn2}
\end{eqnarray}
Since $A_t=B_t$ and $A_\eta= B_\eta$, the WZW-like actions are manifestly equal under this field redefinition. 

Following the discussion of section \ref{subsec:MC}, we know that this field redefinition between must be equivalent to a Maurer-Cartan gauge transformation. Here we should note that Maurer-Cartan gauge transformations do not transform the fields---rather they transform the potentials, and therefore the action, while keeping the string field fixed. But the net effect is equivalent to keeping the action fixed while transforming the string field. With this understanding, the field redefinition between the Berkovits and lifted $A_\infty$ theories is equivalent to a finite Maurer-Cartan gauge transformation
\begin{equation}
B' = \Big[(d+\eta)U\Big] U^{-1}+U B U^{-1},
\end{equation}
where $B$ is the multi-potential of the Berkovits theory and $B'$ is the transformed multi-potential, and the finite gauge parameter $U$ is
\begin{equation}
U(t) =  \overleftarrow{\mathcal{P}}\exp\left[\int_{0}^{t} ds\, A_t[\PhiB](s)\right]e^{-\PhiB(t)},
\end{equation}
where $A_t[\PhiB](t)$ is the $t$-potential of the lifted $A_\infty$ theory evaluated on the Berkovits string field. The net effect of this transformation is to replace the group-like element parameterized by $e^{\PhiB}$ with the group-like element parameterized as $\overleftarrow{\mathcal{P}}\exp\left[\int_{0}^{t} ds\, A_t[\PhiB](s)\right]$. The Berkovits action will then be replaced with the lifted $A_\infty$ action
\begin{equation}
S_B[\PhiB]\ \rightarrow\ S_A[\PhiB],
\end{equation}
after which we may as well rename $\PhiB$ as $\PhiA$. Note that because $U$ only has a zero form component, the Maurer-Cartan gauge transformation does not generate expectation values for the higher potentials. Therefore the transformed multi-potential $B'$ gives a representation of the WZW-like action for the lifted $A_\infty$ theory where all field strengths vanish.

\subsection{From Berkovits to $A_\infty$}
\label{subsec:BA}

So far we have described the field redefinition between the Berkovits and lifted $A_\infty$ theories by equating the $t$-potentials. However, an equivalent characterization can be found by equating the {\it little} $t$-potentials. This naturally leads to a formula for the field redefinition which inverts the Wilson-line of the previous section.

First we need to describe the little potentials of the Berkovits theory. They are defined implicitly by the formula 
\begin{equation}B= \pi_1\G \frac{1}{1-b},\label{eq:Bb}\end{equation}
where the little multi-potential $b$ is a sum of little $n$-potentials 
\begin{equation}b\equiv b^{(0)}+b^{(1)} + b^{(2)} + b^{(3)} + ...,\end{equation}
which can be described using the basis 1-forms $dx^i$ as
\begin{equation}
b^{(0)}=b_\eta,\ \ \ \ \ b^{(1)} = dx^i b_i,\ \ \ \ b^{(2)} = \frac{1}{2!} dx^i\w dx^j\, b_{ij},\ \ \ b^{(3)} = \frac{1}{3!} dx^i\w dx^j\w dx^k\, b_{ijk},\ \ ...\ . \end{equation}
This is precisely analogous to \eq{Aa} of the lifted $A_\infty$ theory. We can invert \eq{Bb} to express $b$ in terms of the multi-potential of the Berkovits theory:
\begin{eqnarray}
\frac{1}{1-B}\lineup = \frac{1}{1-\pi_1\G\frac{1}{1-b}},\nonumber\\
\lineup = \G\frac{1}{1-b}.
\end{eqnarray}
Multiplying this equation by $\G^{-1}$ and projecting onto the 1-string component of the tensor algebra gives
\begin{equation}
b=\pi_1\G^{-1}\frac{1}{1-B}.\label{eq:bB}
\end{equation}
Important special cases are
\begin{eqnarray}
b_\eta \lineup = \pi_1\G^{-1}\frac{1}{1-B_\eta},\\
b_t \lineup = \pi_1 \G^{-1} \frac{1}{1-B_\eta}\otimes B_t\otimes\frac{1}{1-B_\eta},\\
b_\delta \lineup = \pi_1 \G^{-1} \frac{1}{1-B_\eta}\otimes B_\delta\otimes\frac{1}{1-B_\eta},\label{eq:bdelta}\\
b_Q \lineup = \pi_1 \G^{-1} \frac{1}{1-B_\eta}\otimes B_Q\otimes\frac{1}{1-B_\eta}.
\end{eqnarray}
Note that the higher little potentials of the Berkovits theory do not vanish, even though the higher potentials do. The little multi-potential satisfies
\begin{equation}
\pi_1(\n +\D)\frac{1}{1-b}=0,\label{eq:littlebMC}
\end{equation}
by analogy with \eq{littleMC}. 

Next let us explain how to express the Berkovits action in terms of little potentials. Consider the integrand
\begin{eqnarray}
-\langle B_t, Q B_\eta\rangle \lineup = \omega_L(B_t,QB_\eta),\nonumber\\
\lineup = \omega_L\left(\pi_1{\bf B}_t\frac{1}{1-B_\eta},\pi_1\Q\frac{1}{1-B_\eta}\right) ,
\end{eqnarray}
where ${\bf B}_t$ is the coderivation corresponding to $B_t$ regarded as a zero-string product. Since the cohomomorphism $\G^{-1}$ is cyclic, we can use \eq{Id1} to write
\begin{eqnarray}
-\langle B_t, Q B_\eta\rangle \lineup = \omega_L\left(\pi_1\G^{-1} {\bf B}_t\frac{1}{1-B_\eta},\pi_1\G^{-1}\Q\frac{1}{1-B_\eta}\right),\nonumber\\
\lineup = \omega_L\left(\pi_1\G^{-1}\frac{1}{1-B_\eta}\otimes B_t\otimes \frac{1}{1-B_\eta},\pi_1\M\G^{-1}\frac{1}{1-B_\eta}\right),\nonumber\\
\lineup = \omega_L\left(\pi_1\G^{-1}\frac{1}{1-B_\eta}\otimes B_t\otimes \frac{1}{1-B_\eta},\pi_1\M\frac{1}{1-\pi_1\G^{-1}\frac{1}{1-B_\eta}}\right),
\nonumber\\
\lineup = \omega_L\left(b_t,\pi_1\M \frac{1}{1-b_\eta}\right),
\end{eqnarray}
where we substituted the definition of the little potentials. Next we use \eq{littlebMC} to note that
\begin{equation}
\pi_1\M\frac{1}{1-b_\eta} = -\eta b_Q.
\end{equation}
Integrating $t$ from $0$ to $1$ therefore expresses the Berkovits action in the form
\begin{equation}
S_B = -\int_0^1 dt\, \omega_L(b_t,\eta b_Q).
\end{equation}
This is the Berkovits action expressed in terms of little potentials.

\begin{figure}
\begin{center}
\setlength{\unitlength}{.25cm}
\begin{picture}(24,16)
\put(12.875,-.5){$\PhiB$}
\put(12.875,15.5){$\PhiA$}
\put(0,7.5){$\PhiB(t)$}
\put(24,7.5){$\PhiA(t)$}
\put(12.5,1.25){\vector(-3,2){8.5}}
\put(15,1.25){\vector(3,2){8.5}}
\put(12.5,14.75){\vector(-3,-2){8.5}}
\put(15,14.75){\vector(3,-2){8.5}}
\put(13.5,1.5){\vector(0,1){13}}
\put(14.25,14.25){\vector(0,-1){13}}
\put(10,7.5){$\scriptstyle{T(1,0)}$}
\put(14.75,7.5){$\scriptstyle{g(1,0)}$}
\put(19.5,2.75){$\scriptstyle{T(t,0)}$}
\put(19,12.75){$\scriptstyle{f_A(t,\PhiA)}$}
\put(5.75,12.75){$\scriptstyle{g(t,0)}$}
\put(3.75,2.75){$\scriptstyle{f_B(t,\PhiB)}$}
\end{picture}
\end{center}
\caption{\label{fig:FR} A diagram showing the maps which relate the fields and interpolations of the Berkovits and lifted $A_\infty$ superstring field theories.}
\end{figure}
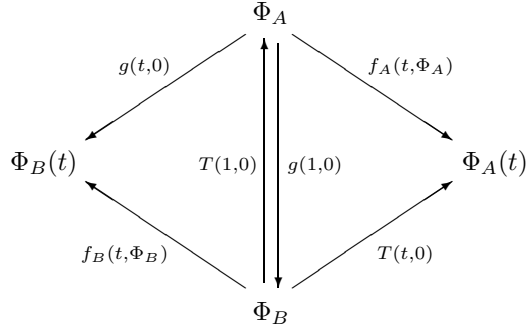

Now it turns out that equating the $t$-potentials of the Berkovits and lifted $A_\infty$ theories is equivalent to equating the little $t$ potentials. To see this, note that 
\begin{equation}
\frac{1}{1-A_\eta}\otimes A_t\otimes\frac{1}{1-A_\eta} = \frac{1}{1-B_\eta}\otimes B_t\otimes\frac{1}{1-B_\eta}.
\end{equation}
where we used the fact that $A_t=B_t$ implies $A_\eta=B_\eta$. The left hand side can be expanded as follows
\begin{eqnarray}
\frac{1}{1-A_\eta}\otimes A_t\otimes\frac{1}{1-A_\eta} \lineup = \frac{1}{1-\pi_1\G\frac{1}{1-\eta\PhiA(t)}}\otimes\left( \pi_1\G\frac{1}{1-\eta\PhiA(t)}\otimes\dot{\Phi}_A(t)\otimes\frac{1}{1-\PhiA(t)}\right)\otimes\frac{1}{1-\pi_1\G\frac{1}{1-\eta\PhiA(t)}},\ \ \ \ \ \nonumber\\
\lineup = \G \frac{1}{1-\eta\PhiA(t)}\otimes\dot{\Phi}_A(t)\otimes\frac{1}{1-\PhiA(t)},\nonumber\\
\lineup = \G \frac{1}{1-a_\eta}\otimes a_t\otimes\frac{1}{1-a_\eta}.
\end{eqnarray}
Therefore
\begin{equation}
\G \frac{1}{1-a_\eta}\otimes a_t\otimes\frac{1}{1-a_\eta} = \frac{1}{1-B_\eta}\otimes B_t\otimes\frac{1}{1-B_\eta}.
\end{equation}
Multiplying this equation by $\G^{-1}$ and projecting onto the 1-string component then implies
\begin{equation}a_t = b_t.\end{equation}
Since $a_t(t) = \dot{\Phi}_A(t)$ this equation is easily integrated to express $\PhiA$ in terms of $\PhiB$.
The solution of this equation is defined by the integral 
\begin{equation}T(t_2,t_1)\equiv \int_{t_1}^{t_2} ds\, b_t(s).\end{equation}
The interpolating field of the lifted $A_\infty$ theory is therefore,
\begin{equation}\PhiA(t) = T(t,0)=\int_0^t ds\, b_t(s),\label{eq:ABt}\end{equation}
and the field redefinition from the Berkovits to the lifted $A_\infty$ theory is 
\begin{equation}\PhiA =T(1,0)= \int_0^1 ds\, b_t(s).\label{eq:AB}\end{equation}
Similar to the previous section (but in reverse), this expresses $\PhiA$ as a function of an entire path $\PhiB(t)$ in the Berkovits theory. For this to be a field redefinition between $\PhiA$ and $\PhiB$, we assume that $\PhiB(t)$ is specified by an interpolating function $f_B(t,\PhiB)$ whose zero-string product vanishes. Equation \eq{ABt} determines the interpolating function $f_A(t,\PhiA)$ of the lifted $A_\infty$ theory in terms of the interpolating function $f_B(\PhiB,t)$ of the Berkovits theory. We summarize the different fields, interpolations, and mappings between them in figure \ref{fig:FR}.

We should emphasize that this is the same as the Wilson line field redefinition of the previous section, but inverted. To see this, suppose we express the Berkovits interpolation $\PhiB(t)$ in terms of the lifted $A_\infty$ interpolation $\PhiA(t)$ using the Wilson line \eq{BAt}. This allows us to write $B_t=A_t$ and $B_\eta=A_\eta$, and substituting into \eq{ABt} gives 
\begin{eqnarray}
\PhiA(t) \lineup = \int_0^t ds\, \pi_1 \G^{-1}\frac{1}{1-A_\eta(s)}\otimes A_t(s)\otimes\frac{1}{1-A_\eta(s)},\nonumber\\
\lineup = \int_0^t ds\, \pi_1 \G^{-1}\G\frac{1}{1-\eta\PhiA(s)}\otimes \dot{\Phi}_A(s)\otimes\frac{1}{1-\eta\PhiA(s)},\nonumber\\
\lineup = \int_0^t ds\, \dot{\Phi}_A(s),\nonumber\\
\lineup = \PhiA(t),
\end{eqnarray}
which is the string field we started with.

\subsection{Okawa's Approach}

Let us describe an unrelated proposal for the field redefinition suggested by Y. Okawa.\footnote{The author would like to thank him for sharing this idea.} This approach does not consider the interpolations $\PhiA(t)$ and $\PhiB(t)$, but focuses on the dynamical fields $\PhiA$ and $\PhiB$ at $t=1$. The starting point is the condition
\begin{equation}B_\eta|_{t=1}=A_\eta|_{t=1}.\label{eq:AnBn3}\end{equation}
As mentioned before, this relation does not fully constrain the field redefinition. However, this can be remedied with a few additional choices. Equating the 
$\eta$-potentials is equivalent to equating the little $\eta$-potentials:
\begin{equation}
a_\eta|_{t=1} = b_\eta|_{t=1}.
\end{equation}
Since $a_\eta|_{t=1} = \eta\PhiA$, this gives 
\begin{equation}
\eta\PhiA = b_\eta|_{t=1}.\label{eq:Ok1}
\end{equation}
Suppose we assume that the lifted $A_\infty$ field satisfies the gauge condition $\xi\PhiA = 0$. Then the above relation implies that the lifted $A_\infty$ field satisfies
\begin{equation}
\PhiA = \xi b_\eta|_{t=1} \ \ \ \ \ \ \mathrm{if}\ \xi\PhiA = 0.
\end{equation}
This is not quite a field redefinition between $\PhiA$ and $\PhiB$ since it is not invertible---the field $\PhiA$ always satisfies the gauge condition $\xi\PhiA=0$. We can fix this by adding the $\eta$-closed term $\eta\xi\PhiB$, so the proposed field redefinition is 
\begin{equation}\PhiA = \eta\xi\PhiB+\xi b_\eta|_{t=1}.\label{eq:Yuji} \end{equation}
Taking $\eta$ of this expression implies $a_\eta|_{t=1}=b_\eta|_{t=1}$, as desired. A nice property of this field redefinition is that it is compatible with the natural gauge fixing to the small Hilbert space. In particular, fixing $\xi\PhiA = 0$ in the lifted $A_\infty$ theory fixes $\xi\PhiB=0$ in the Berkovits theory, and vice-versa:
\begin{equation}\xi\PhiA= 0\ \longleftrightarrow\ \xi\PhiB = 0.\end{equation}
This is not generally true for the field redefinition based on the Wilson line.

Expanding the field redefinition to second order gives
\begin{equation}\PhiB = \PhiA+\frac{1}{3}\xi[\xi\eta\PhiA,\eta\PhiA]-\frac{1}{2}\xi[\PhiA,\eta\PhiA]+\mathrm{higher\ orders}.\end{equation}
Note that this is {\it not} a special case of \eq{BA2nd} for some choice of the constant $C$. Therefore the field redefinition cannot be realized as the endpoint of a pair of interpolations $\PhiA(t)$ and $\PhiB(t)$ related by $A_t=B_t$. This makes the proof of equivalence of the actions less direct. First let us generalize the field redefinition to intermediate $t$ by taking
\begin{equation}
\PhiA(t) = \eta\xi\PhiB(t)+\xi b_\eta(t).\label{eq:Yujit}
\end{equation}
With this identification the $\eta$-potentials are equal for intermediate $t$
\begin{equation}
A_\eta(t) = B_\eta(t),
\end{equation}
but the $t$-potentials are not the same. Instead we have 
\begin{equation}A_t = \tilde{B}_t,
\end{equation}
where $\tilde{B}_t$ is defined
\begin{eqnarray}
\tilde{B}_t \lineup \equiv \pi_1 \G\frac{1}{1-b_\eta}\otimes\Big(\eta\xi\dot{\Phi}_B(t)+\xi\dot{b}_\eta \Big)\otimes \frac{1}{1-b_\eta},\label{eq:YujiAt}
\end{eqnarray}
which is not the same as $B_t$. Therefore applying the field redefinition to the lifted $A_\infty$ action gives 
\begin{equation}
S_B= -\int_0^1 dt\,\langle \tilde{B}_t, QB_\eta\rangle_L.\label{eq:YujiAction}
\end{equation}
This is not the Berkovits action as it is usually written. However, as noted in \cite{OkWB}, $\tilde{B}_t$ differs from $B_t$ by a term which is annihilated by $\nabla_\eta$, so in fact \eq{YujiAction} is the same as the usual Berkovits action. Another way to understand this observation is that \eq{YujiAction} is an expression of the Berkovits action using a nonstandard set of potentials, related to the usual Berkovits potentials by a finite Maurer-Cartan gauge transformation
\begin{equation}
\tilde{B} =\Big[(d+\eta)e^{\Lambda^{(1)}}\Big] e^{-\Lambda^{(1)}}+ e^{\Lambda^{(1)}} B e^{-\Lambda^{(1)}},
\end{equation}
where the 1-form gauge parameter $\Lambda^{(1)}$ is
\begin{equation}
\Lambda^{(1)} = dt \Lambda_t.
\end{equation}
This Maurer-Cartan gauge transformation leaves $B_\eta$ invariant, while it transforms $B_t$ into $\tilde{B}_t$:
\begin{equation}
\tilde{B}_t=B_t+\nabla_\eta \Lambda_t.
\end{equation}
We can compute the gauge parameter $\Lambda_t$ as follows. Consider
\begin{eqnarray}
\tilde{B}_t-B_t \lineup = \pi_1 \G\frac{1}{1-b_\eta}\otimes\Big(\eta\xi\dot{\Phi}_B(t)+\xi\dot{b}_\eta \Big)\otimes \frac{1}{1-b_\eta} - B_t,\nonumber\\
\lineup = \pi_1 \G\frac{1}{1-b_\eta}\otimes\Big(\eta\xi\dot{\Phi}_B(t)+\xi\eta b_t \Big)\otimes \frac{1}{1-b_\eta} - B_t,
\end{eqnarray}
where in the second step we traded $\dot{b}_\eta$ with $\eta b_t$. Now pull $\eta$ out so it acts on the entire expression:
\begin{eqnarray}
\tilde{B}_t- B_t\lineup = \pi_1 \G\frac{1}{1-b_\eta}\otimes\Big(\eta\xi(\dot{\Phi}_B(t)-b_t)+ b_t \Big)\otimes \frac{1}{1-b_\eta} - B_t,\nonumber\\
\lineup = \pi_1 \G\n \frac{1}{1-b_\eta}\otimes \xi(\dot{\Phi}_B(t)-b_t) \otimes \frac{1}{1-b_\eta} +\pi_1 \G \frac{1}{1-b_\eta}\otimes b_t \otimes \frac{1}{1-b_\eta} - B_t,\nonumber\\
\lineup = \pi_1(\n-\m_2) \G\frac{1}{1-b_\eta}\otimes \xi(\dot{\Phi}_B(t)-b_t) \otimes \frac{1}{1-b_\eta}.
\end{eqnarray}
In the second step the second term cancels with $B_t$ by the definition of the little potentials. What remains can be interpreted as $\nabla_\eta\Lambda_t$,
where $\Lambda_t$ is 
\begin{equation}
\Lambda_t = \pi_1 \G\frac{1}{1-b_\eta}\otimes \xi(\dot{\Phi}_B(t)-b_t) \otimes \frac{1}{1-b_\eta}.
\end{equation}
Therefore relating the interpolations through \eq{Yujit} and performing a Maurer-Cartan gauge transformation turns the WZW-like action of the lifted $A_\infty$ theory into the standard WZW-like action of the Berkovits theory. 

It is interesting to contrast the variety of field redefinitions we find in the large Hilbert space with the seeming uniqueness of the field redefinition found in the small Hilbert space \cite{OkWB,WB}. The reason for this discrepancy is that the operations $\eta$, $\xi,m_2$ used to construct the field redefinition in the large Hilbert space can also implement $\eta$ gauge transformations of the field redefinition. In the small Hilbert space the $\eta$ gauge invariance is not present, and $\xi$ and $m_2$ alone cannot implement interesting gauge transformations in the small Hilbert space.

\section{Mapping Gauge Invariances}
\label{sec:gauge}

We will now use the field redefinition to determine how the gauge symmetries of the lifted $A_\infty$ theory map into the Berkovits theory, and vice versa.

Let us first consider the Wilson line field redefinition \eq{BA}. To get the information we're after, we must compute the change of the Berkovits field $\PhiB$ induced by a change in the lifted $A_\infty$ field $\PhiA$ and/or the interpolating function $f_A(t,\PhiA)$. Taking the variation of \eq{BA} produces
\begin{equation}\delta e^{\PhiB} = \int_0^1 dt\, g(1,t)\delta A_t\, g(t,0).\end{equation}
Using \eq{dswitch} we can switch the variation with a time derivative:
\begin{eqnarray}\delta e^{\PhiB}\lineup = \int_0^1 dt\, g(1,t)\Big( \nabla_t A_\delta + F_{\delta t}\Big) g(t,0),\nonumber\\
\lineup = \int_0^1 dt\, \frac{d}{dt}\Big( g(1,t) A_\delta g(t,0) \Big) +  \int_0^1 dt\, g(1,t) F_{\delta t}\, g(t,0), \nonumber\\
\lineup = A_\delta|_{t=1} e^{\PhiB}\, -e^{\PhiB}A_\delta|_{t=0} + \, \int_0^1 dt\, g(1,t) F_{\delta t}\, g(t,0).
\end{eqnarray}
This is the expected formula for the variation of a Wilson line. We assume that $A_\delta$ vanishes at $t=0$ because the interpolating function is required to satisfy the boundary condition $f_A(0,\PhiA)=0$. Let us see what to do with the field strength integrated along the curve. We can express the field strength in terms of the 2-potential 
\begin{eqnarray}
\int_0^1 dt\, g(1,t) F_{\delta t}\, g(t,0)  \lineup = -\int_0^1 dt\, g(1,t)\big(\nabla_\eta A_{t\delta}\big) g(t,0) ,\nonumber\\
\lineup =- e^{\PhiB} \int_0^1 dt\, g(t,0)^{-1}\big( \nabla_\eta A_{t\delta}\big) g(t,0),\\
 \lineup = - e^{\PhiB} \int_0^1 dt\, \eta\Big(g(t,0)^{-1}  A_{t\delta} \, g(t,0)\Big).
\end{eqnarray}
Therefore, the variation of the field redefinition is
\begin{equation}
\delta e^{\PhiB} = A_\delta|_{t=1} e^{\PhiB}\, -\, e^{\PhiB}\eta\left( \int_0^1 dt\, g(t,0)^{-1}  A_{t\delta} \, g(t,0)\right),
\label{eq:deB}
\end{equation}
Consider specifically variations of the interpolating function $f_A(t,\PhiA)$. In this case $A_\delta|_{t=1}$ is constrained to vanish by boundary conditions, and  \eq{deB} simplifies to
\begin{equation}
\delta e^{\PhiB} =  - e^{\PhiB}\eta\left(\int_0^1 dt\, g(t,0)^{-1}  A_{t\delta} \, g(t,0)\right).
\end{equation}
Right multiplication of $e^{\PhiB}$ by an $\eta$-closed string field is an infinitesimal gauge transformation in the Berkovits theory. Therefore, a change of the interpolation only effects the field redefinition between $\PhiA$ and $\PhiB$ by an $\eta$ gauge transformation. This is consistent with what we found in \eq{BAgauge}.

Now let us see how $\PhiB$ responds to gauge transformations in the lifted $A_\infty$ theory. First consider the $\eta$ gauge transformation $\delta_\eta$ in \eq{deltaeta}. We obtain
\begin{equation}
\delta_\eta e^{\PhiB} = A_{\delta_\eta}|_{t=1} e^{\PhiB}\, -\, e^{\PhiB}\int_0^1 dt\, \eta\Big(g(t,0)^{-1}  A_{t\delta_\eta} \, g(t,0)\Big).
\end{equation}
We can simplify the first term using equation \eq{OmB}:
\begin{eqnarray}
A_{\delta_\eta}|_{t=1}\lineup = \nabla_\eta\Omega_B|_{t=1},\nonumber\\
\lineup = g(1,0)\eta\Big(g(1,0)^{-1} \Omega_B \, g(1,0)\Big) g(1,0)^{-1}.
\end{eqnarray}
Plugging in we obtain 
\begin{equation}
\delta_\eta e^{\PhiB} = e^{\PhiB}\,\eta\left( g(1,0)^{-1} \Omega_B \, g(1,0) - \int_0^1 dt\, g(t,0)^{-1}  A_{t\delta_\eta} \, g(t,0)\right).
\end{equation}
Therefore, the $\eta$ gauge invariance of the lifted $A_\infty$ theory maps into the $\eta$ gauge invariance of the Berkovits theory. Now consider the BRST gauge transformation $\delta_Q$ in \eq{deltaQ}. From the computation of \eq{AdQ}, we find 
\begin{equation}
\delta_Q e^{\PhiB} = \Big(Q\Lambda_B -\mu_B\Big)e^{\PhiB}\, -\, e^{\PhiB}\eta\left(\int_0^1 dt\, g(t,0)^{-1}  A_{t\delta_Q} \, g(t,0)\right).
\end{equation}
Left multiplication of $e^{\PhiB}$ by a BRST closed string field is an infinitesimal gauge transformation in the Berkovits theory. It follows from the computation at the end of subsection \ref{subsec:var} that left multiplication of $e^{\PhiB}$ by $\mu_B$ is also a symmetry of the action, even though $\mu_B$ is not BRST closed. However, note from \eq{DeltaT} that $\mu_B$ vanishes when the equations of motion are satisfied:
\begin{equation}\pi_1\M\frac{1}{1-\eta\PhiA} = 0.\end{equation}
Therefore left multiplication  by $\mu_B$ must represent a trivial gauge transformation \cite{HT}. The upshot is that the BRST gauge transformation of the lifted $A_\infty$ theory maps into a combination of a BRST gauge transformation, an $\eta$ gauge transformation, and a trivial gauge transformation in the Berkovits theory. 

Now let's consider the reverse question, namely, how the gauge invariances of the Berkovits theory map into those of the lifted $A_\infty$ theory. For this purpose it is useful to consider the inverse field redefinition as described in section \ref{subsec:BA}. Taking the variation of \eq{AB}, one finds that a change of the Berkovits field and/or interpolation changes the lifted $A_\infty$ field through
\begin{eqnarray}
\delta \PhiA\lineup  = \int_0^1 dt\, \delta b_t,\nonumber\\
\lineup = \int_0^1 dt \left(\frac{d}{dt} b_\delta -\eta b_{\delta t}\right),\nonumber\\
\lineup = b_\delta|_{t=1} - \eta\left(\int_0^1 dt\, b_{\delta t}\right),\label{eq:Avar}
\end{eqnarray}
where we used \eq{littlebMC} to interchange $\delta$ with a $d/dt$, which produces a term proportional to the little 2-potential. If we change the interpolating function of the Berkovits theory, the boundary term at $t=1$ drops out and what remains is an $\eta$ gauge transformation of $\PhiA$. The BRST and $\eta$ gauge transformations of the Berkovits theory can be written
\begin{equation}\delta_Q e^{\PhiB} = \big(Q\Lambda_B \big)e^{\PhiB},\ \ \ \ \delta_\eta e^{\PhiB} = e^{\PhiB}\eta\Big(e^{-\PhiB}\Omega_B e^{\PhiB}\Big),\end{equation}
Before $\Lambda_B$ and $\Omega_B$ were defined as functions of $\PhiA$ and the gauge parameters $\Lambda_A$ and $\Omega_A$ of the lifted $A_\infty$ theory, but now we view them as independent variables defining the gauge parameters of the Berkovits theory.  The potentials corresponding to these variations are 
\begin{equation}B_{\delta_Q}|_{t=1} = Q\Lambda_B,\ \ \ \ B_{\delta_\eta}|_{t=1} = \nabla_\eta \Omega_B|_{t=1}.\end{equation}
Let us first compute the little potential $b_{\delta_\eta}|_{t=1}$ from \eq{bdelta}:
\begin{eqnarray}
b_{\delta_\eta}|_{t=1} \lineup = \left.\pi_1 \G^{-1} \frac{1}{1-B_\eta}\otimes B_{\delta_\eta}\otimes\frac{1}{1-B_\eta}\right|_{t=1},\nonumber\\
\lineup = \left.\pi_1 \G^{-1} \frac{1}{1-B_\eta}\otimes \nabla_\eta \Omega_B\otimes\frac{1}{1-B_\eta}\right|_{t=1},\nonumber\\
\lineup = \left.\pi_1 \G^{-1}(\n-\m_2) \frac{1}{1-B_\eta}\otimes \Omega_B\otimes\frac{1}{1-B_\eta}\right|_{t=1},\nonumber\\
\lineup = \eta \left.\pi_1 \G^{-1} \frac{1}{1-B_\eta}\otimes \Omega_B\otimes\frac{1}{1-B_\eta}\right|_{t=1}.
\end{eqnarray}
Plugging this into \eq{Avar}, we find 
\begin{equation}
\delta_\eta \PhiA = \eta\left(\Omega_A- \int_0^1 dt\, b_{\delta_\eta t}\right),
\end{equation}
where
\begin{equation}
\Omega_A \equiv \left.\pi_1\G^{-1} \frac{1}{1-B_\eta}\otimes \Omega_B\otimes\frac{1}{1-B_\eta}\right|_{t=1}.\label{eq:OmA}
\end{equation}
Therefore the $\eta$ gauge invariance of the Berkovits theory maps into the $\eta$ gauge invariance of the lifted $A_\infty$ theory. Note that \eq{OmA} is the inverse of the formula \eq{OmB} expressing $\Omega_B$ as a function of $\Omega_A$. Therefore we have a ``field redefinition" relating the $\eta$ gauge parameters in the two theories. Now consider the BRST gauge symmetry:
\begin{eqnarray}
b_{\delta_Q} \lineup = \left.\pi_1 \G^{-1} \frac{1}{1-B_\eta}\otimes B_{\delta_Q}\otimes\frac{1}{1-B_\eta}\right|_{t=1},\nonumber\\
\lineup = \left.\pi_1 \G^{-1} \frac{1}{1-B_\eta}\otimes Q\Lambda_B\otimes\frac{1}{1-B_\eta}\right|_{t=1},\nonumber\\
\lineup = \left.\pi_1 \G^{-1} \Q \frac{1}{1-B_\eta}\otimes \Lambda_B\otimes\frac{1}{1-B_\eta}\right|_{t=1}\nonumber\\
\lineup\ \ \  - \left.\pi_1 \G^{-1}\left( \frac{1}{1-B_\eta}\otimes QB_\eta\otimes \frac{1}{1-B_\eta}\otimes \Lambda_B\otimes\frac{1}{1-B_\eta}- \frac{1}{1-B_\eta}\otimes \Lambda_B\otimes\frac{1}{1-B_\eta}\otimes QB_\eta\otimes \frac{1}{1-B_\eta}\right)\right|_{t=1},\nonumber\\
\lineup = \left.\pi_1 \M\G^{-1} \frac{1}{1-B_\eta}\otimes \Lambda_B\otimes\frac{1}{1-B_\eta}\right|_{t=1}\nonumber\\
\lineup\ \ \  - \left.\pi_1 \G^{-1}\left( \frac{1}{1-B_\eta}\otimes QB_\eta\otimes \frac{1}{1-B_\eta}\otimes \Lambda_B\otimes\frac{1}{1-B_\eta}- \frac{1}{1-B_\eta}\otimes \Lambda_B\otimes\frac{1}{1-B_\eta}\otimes QB_\eta\otimes \frac{1}{1-B_\eta}\right)\right|_{t=1},\nonumber\\
\lineup = \left.\pi_1\M\frac{1}{1-\pi_1\G^{-1}\frac{1}{1-B_\eta}}\otimes\left(\pi_1 \G^{-1}\frac{1}{1-B_\eta}\otimes \Lambda_B\otimes\frac{1}{1-B_\eta}\right)\otimes \frac{1}{1-\pi_1\G^{-1}\frac{1}{1-B_\eta}}\right|_{t=1}\nonumber\\
\lineup\ \ \  - \left.\pi_1 \G^{-1}\left( \frac{1}{1-B_\eta}\otimes QB_\eta\otimes \frac{1}{1-B_\eta}\otimes \Lambda_B\otimes\frac{1}{1-B_\eta}- \frac{1}{1-B_\eta}\otimes \Lambda_B\otimes\frac{1}{1-B_\eta}\otimes QB_\eta\otimes \frac{1}{1-B_\eta}\right)\right|_{t=1}\nonumber\\
\end{eqnarray}
Using 
\begin{equation}
\eta\PhiA = \left.\pi_1\G^{-1}\frac{1}{1-B_\eta}\right|_{t=1},
\end{equation}
we therefore obtain
\begin{equation}
\delta_Q\PhiA = \left(\pi_1\M\frac{1}{1-\eta\PhiA}\otimes\Lambda_A\otimes \frac{1}{1-\eta\PhiA}\right)-\mu_A - \eta\left(\int_0^1 dt\, b_{\delta_Q t}\right),
\end{equation}
where 
\begin{eqnarray}
\Lambda_A\lineup \equiv\left.\pi_1 \G^{-1}\frac{1}{1-B_\eta}\otimes \Lambda_B\otimes\frac{1}{1-B_\eta}\right|_{t=1},\label{eq:LambdaA}\\
\mu_A \lineup \equiv \left.\pi_1 \G^{-1}\left( \frac{1}{1-B_\eta}\otimes QB_\eta\otimes \frac{1}{1-B_\eta}\otimes \Lambda_B\otimes\frac{1}{1-B_\eta}+ \frac{1}{1-B_\eta}\otimes \Lambda_B\otimes\frac{1}{1-B_\eta}\otimes QB_\eta\otimes \frac{1}{1-B_\eta}\right)\right|_{t=1}.\ \ \ \ \ \ \ \ 
\end{eqnarray}
Note that \eq{LambdaA} is the inverse of the formula \eq{LambdaB} expressing $\Lambda_B$ in terms of $\Lambda_A$. Also note that $\mu_A$ vanishes on shell, and so must represent a trivial gauge transformation. Therefore the BRST gauge transformation of the Berkovits theory maps into a combination of a BRST, an $\eta$, and trivial gauge transformations in the lifted $A_\infty$ theory. Similar conclusions follow using the field redefinition proposed by Okawa, since in this case the variation takes the form
\begin{equation}
\delta\PhiA =\left. b_\delta\right|_{t=1} + \eta\xi\Big(\delta\PhiB -\left.b_\delta\right|_{t=1}\Big).
\end{equation}
which, aside from unimportant differences in the $\eta$ closed term, is equivalent to \eq{Avar}.

\vspace{.5cm}

\noindent{\bf Acknowledgments}

\vspace{.25cm}

\noindent The author would like to thank T. Takezaki and Y. Okawa for collaboration, and S. Konopka and Y. Okawa for comments. This work was supported in parts by the DFG Transregional Collaborative Research Centre TRR 33 and the DFG cluster of excellence Origin and Structure of the Universe.

\begin{appendix}

\section{Some Computations Involving Cyclicity}
\label{app:cyclic}

In this appendix we provide a few missing calculations referred to in the text, in particular as pertains to cyclicity of the $A_\infty$ products and cohomomorphism $\G$. These calculations are simplified with the help of the ``triangle formalism" of the product and coproduct, introduced in appendix A of \cite{WB}. Here we review this formalism and provide the missing calculations in the text.

The tensor algebra has a {\it coproduct}, which is a coassociative linear map from the tensor algebra into a {\it pair} of tensor algebras:
\begin{equation}\triangle: T\mathcal{H}\to T\mathcal{H}\otimes'T\mathcal{H},\end{equation}
where we use the symbol $\otimes'$ to distinguish from the tensor product used to construct $T\mathcal{H}$. The coproduct is {\it coassociative}
\begin{equation}
(\triangle \otimes' \mathbb{I}_{T\mathcal{H}})\triangle = (\mathbb{I}\otimes' \triangle)\triangle,
\end{equation}
and acts on tensor products of states as 
\begin{equation}
\triangle A_1\otimes ... \otimes A_n = \sum_{k=0}^n (A_1\otimes ...\otimes A_k) \otimes' (A_{k+1}\otimes...\otimes A_n),
\end{equation}
where at the extremes of summation $\otimes'$ multiplies the identity of the tensor product $1_{T\mathcal{H}}$. Note that $1_{T\mathcal{H}}$ is not the identity with respect to $\otimes'$. Coderivations and cohomomorphisms satisfy
\begin{eqnarray}
\triangle \D \lineup = (\D\otimes'\mathbb{I}_{T\mathcal{H}}+\mathbb{I}_{T\mathcal{H}}\otimes' \D)\triangle,\\
\triangle\H \lineup = (\H\otimes'\H)\triangle,
\end{eqnarray}
and group-like elements satisfy
\begin{equation}\triangle\frac{1}{1-A} = \frac{1}{1-A}\otimes'\frac{1}{1-A}.\end{equation}
By taking variations we can derive the action of the coproduct on more general states. In addition, the tensor algebra has a product
\begin{equation}\ \inverttriangle :T\mathcal{H}\otimes'T\mathcal{H}\to T\mathcal{H},\end{equation}
which operates by replacing the primed tensor product $\otimes'$ with the ordinary tensor product $\otimes$. The central formula of the triangle formalism is an expression for the projector $\pi_{m+n}$ onto the $(m+n)$-string component of the tensor algebra:
\begin{equation}\pi_{m+n} = \inverttriangle\!\Big[\pi_m\otimes'\pi_n\Big]\triangle.\end{equation}
For further elaboration, see appendix A of \cite{WB}.

\subsection{Proof of \eq{Id1}}

Now let us revisit the derivation of \eq{Id1}, which is also featured in appendix A of \cite{OkWB}. A cyclic cohomomorphism $\H$ satisfies 
\begin{equation}\langle \omega|\pi_2\H = \langle \omega|\pi_2.\end{equation}
Consider this formula acting on a particular element of the tensor algebra: 
\begin{equation}\frac{1}{1-A}\otimes B\otimes \frac{1}{1-A}\otimes C\otimes\frac{1}{1-A},\end{equation}
where $A$ is degree even and $B,C$ are arbitrary string fields. We find
\begin{eqnarray}\langle \omega |\pi_2 \H \left(\frac{1}{1-A}\otimes B\otimes \frac{1}{1-A}\otimes C\otimes\frac{1}{1-A}\right)\lineup =
\langle \omega |\pi_2 \left(\frac{1}{1-A}\otimes B\otimes \frac{1}{1-A}\otimes C\otimes\frac{1}{1-A}\right),\nonumber\\
\lineup = \langle \omega | B\otimes C.
\label{eq:A7}\end{eqnarray}
On the other hand, we can replace $\pi_2=\inverttriangle\!\Big[\pi_1\otimes'\pi_1\Big]\triangle$ on the left hand side and act with the coproduct:
\begin{eqnarray}
\lineup \langle \omega |\pi_2 \H \left(\frac{1}{1-A}\otimes B\otimes \frac{1}{1-A}\otimes C\otimes\frac{1}{1-A}\right) =\langle \omega |\inverttriangle\!\Big[\pi_1\otimes'\pi_1\Big]\triangle \H \left(\frac{1}{1-A}\otimes B\otimes \frac{1}{1-A}\otimes C\otimes\frac{1}{1-A}\right),\nonumber\\
\lineup\ \ \ =\langle \omega |\inverttriangle\!\Bigg[(\pi_1\otimes'\pi_1)(\H\otimes'\H)\Bigg]\triangle\left( \frac{1}{1-A}\otimes B\otimes \frac{1}{1-A}\otimes C\otimes\frac{1}{1-A}\right),\nonumber\\
\lineup\ \ \ =\langle \omega |\inverttriangle\!\Bigg[(\pi_1\otimes'\pi_1)(\H\otimes'\H)\left( \frac{1}{1-A}\otimes'\frac{1}{1-A}\otimes B\otimes \frac{1}{1-A}\otimes C\otimes\frac{1}{1-A}\right.\nonumber\\
\lineup\ \ \ \ \ \ \ \ \ \ \ \left.+\frac{1}{1-A}\otimes B\otimes \frac{1}{1-A}\otimes'\frac{1}{1-A}\otimes C\otimes\frac{1}{1-A}+\frac{1}{1-A}\otimes B\otimes \frac{1}{1-A}\otimes C\otimes\frac{1}{1-A}\otimes'\frac{1}{1-A}\right)\Bigg],\nonumber\\
\lineup = \langle \omega |\inverttriangle\! \Bigg[\left(\pi_1\H\frac{1}{1-A}\otimes B\otimes \frac{1}{1-A}\otimes C\otimes\frac{1}{1-A}\right)\otimes' \left(\pi_1\H\frac{1}{1-A}\right)\nonumber\\
\lineup\ \ \ \ \ \ \ \ \ \ \ \ + \left(\pi_1\H\frac{1}{1-A}\otimes B\otimes \frac{1}{1-A}\right)\otimes' \left(\pi_1\H\frac{1}{1-A}\otimes C\otimes\frac{1}{1-A}\right)\nonumber\\
\lineup \ \ \ \ \ \ \ \ \ \ \ \ + \left(\pi_1\H \frac{1}{1-A}\right)\otimes' \left(\pi_1\H \frac{1}{1-A}\otimes B\otimes \frac{1}{1-A}\otimes C\otimes\frac{1}{1-A}\right)\Bigg],\nonumber\\
\lineup = \langle \omega | \left(\pi_1\H\frac{1}{1-A}\otimes B\otimes \frac{1}{1-A}\otimes C\otimes\frac{1}{1-A}\right)\otimes \left(\pi_1\H\frac{1}{1-A}\right)\nonumber\\
\lineup\ \ \ +\langle \omega |\left(\pi_1\H\frac{1}{1-A}\otimes B\otimes \frac{1}{1-A}\right)\otimes \left(\pi_1\H\frac{1}{1-A}\otimes C\otimes\frac{1}{1-A}\right)\nonumber\\
\lineup \ \ \ + \langle \omega |\left(\pi_1\H \frac{1}{1-A}\right)\otimes \left(\pi_1\H \frac{1}{1-A}\otimes B\otimes \frac{1}{1-A}\otimes C\otimes\frac{1}{1-A}\right).
\end{eqnarray}
The first and last terms cancel by antisymmetry of the symplectic form. The second term, however, remains. We have therefore shown
\begin{equation}
\omega\left(\pi_1\H \left(\frac{1}{1-A}\otimes B\otimes \frac{1}{1-A}\right), \pi_1\H\left(\frac{1}{1-A}\otimes C\otimes\frac{1}{1-A}\right)\right) = \omega(A,B).
\end{equation}
Now suppose that the string fields $A$ and $B$ happen to take the form
\begin{equation}
A=\pi_1 \D_1\frac{1}{1-A},\ \ \ B = \pi_1\D_2\frac{1}{1-A},
\end{equation}
for some coderivations $\D_1$ and $\D_2$. Plugging into the above formula then gives 
\begin{equation}
\omega\left(\pi_1\H \D_1\left(\frac{1}{1-A}\right), \pi_1\H\D_2\left(\frac{1}{1-A}\right)\right) = \omega\left(\pi_1\D_1\left(\frac{1}{1-A}\right),\pi_1\D_2\left(\frac{1}{1-A}\right)\right),
\end{equation}
which reproduces \eq{Id1}.

\subsection{Proof of \eq{Id2}}

Now let's prove the equivalence of equations \eq{Id2} and \eq{dQvar}. The left hand side of \eq{Id2} is
\begin{equation}
\langle \omega_L|\pi_2 \G \left[\frac{1}{1-\eta\PhiA}\otimes \left(\pi_1 \M \frac{1}{1-\eta\PhiA}\right)\otimes \frac{1}{1-\eta\PhiA}\otimes\Lambda\otimes\frac{1}{1-\eta\PhiA}\otimes\left(\pi_1\M \frac{1}{1-\eta\PhiA}\right)\otimes  \frac{1}{1-\eta\PhiA}\right].
\end{equation}
Replacing $\pi_2=\inverttriangle\! \Big[\pi_1\otimes'\pi_1\Big]\triangle$ and acting with the coproduct produces the expression 
\begin{eqnarray}
\lineup\langle \omega_L|\inverttriangle\!(\pi_1\otimes'\pi_1)(\G\otimes'\G)\nonumber\\
\lineup\ \ \ \times \left[\frac{1}{1-\eta\PhiA}\otimes'\frac{1}{1-\eta\PhiA}\otimes \left(\pi_1 \M \frac{1}{1-\eta\PhiA}\right)\otimes \frac{1}{1-\eta\PhiA}\otimes\Lambda\otimes\frac{1}{1-\eta\PhiA}\otimes\left(\pi_1\M \frac{1}{1-\eta\PhiA}\right)\otimes  \frac{1}{1-\eta\PhiA}\right.\ \ \ \ \ \ \ \ \ \ 
\nonumber\\
\lineup\ \ \ \ \ +\frac{1}{1-\eta\PhiA}\otimes \left(\pi_1 \M \frac{1}{1-\eta\PhiA}\right)\otimes \frac{1}{1-\eta\PhiA}\otimes'\frac{1}{1-\eta\PhiA}\otimes\Lambda\otimes\frac{1}{1-\eta\PhiA}\otimes\left(\pi_1\M \frac{1}{1-\eta\PhiA}\right)\otimes  \frac{1}{1-\eta\PhiA}\nonumber\\
\lineup\ \ \ \ \ +\frac{1}{1-\eta\PhiA}\otimes \left(\pi_1 \M \frac{1}{1-\eta\PhiA}\right)\otimes \frac{1}{1-\eta\PhiA}\otimes\Lambda\otimes\frac{1}{1-\eta\PhiA}\otimes'\frac{1}{1-\eta\PhiA}\otimes\left(\pi_1\M \frac{1}{1-\eta\PhiA}\right)\otimes  \frac{1}{1-\eta\PhiA}\nonumber\\
\lineup\ \ \ \ \ \left.+\frac{1}{1-\eta\PhiA}\otimes \left(\pi_1 \M \frac{1}{1-\eta\PhiA}\right)\otimes \frac{1}{1-\eta\PhiA}\otimes\Lambda\otimes\frac{1}{1-\eta\PhiA}\otimes\left(\pi_1\M \frac{1}{1-\eta\PhiA}\right)\otimes  \frac{1}{1-\eta\PhiA}\otimes'\frac{1}{1-\eta\PhiA}\right].
\label{eq:stepId2}\end{eqnarray}
The first and last term simplify to
\begin{eqnarray}
\omega_L\left(\!\pi_1\G\frac{1}{1-\eta\PhiA},\pi_1\G \frac{1}{1-\eta\PhiA}\!\otimes\! \left(\pi_1 \M \frac{1}{1-\eta\PhiA}\right)\!\otimes\! \frac{1}{1-\eta\PhiA}\!\otimes\!\Lambda\!\otimes\!\frac{1}{1-\eta\PhiA}\!\otimes\!\left(\pi_1\M \frac{1}{1-\eta\PhiA}\right)\!\otimes\!\frac{1}{1-\eta\PhiA}\right)\ \ \lineup \nonumber\\
+\omega_L\left(\pi_1\G \frac{1}{1-\eta\PhiA}\!\otimes\! \left(\pi_1 \M \frac{1}{1-\eta\PhiA}\right)\!\otimes\! \frac{1}{1-\eta\PhiA}\!\otimes\!\Lambda\!\otimes\!\frac{1}{1-\eta\PhiA}\!\otimes\!\left(\pi_1\M \frac{1}{1-\eta\PhiA}\right)\!\otimes\!\frac{1}{1-\eta\PhiA},\pi_1\G \frac{1}{1-\eta\PhiA}\right),\lineup \nonumber\\
\end{eqnarray}
and they cancel by antisymmetry of the symplectic form. Meanwhile, the second and third terms in \eq{stepId2} produce equation \eq{dQvar}. This fills the missing steps between \eq{dQvar} and \eq{Id2}.

\subsection{Proof of \eq{Id3}}

Now let's prove \eq{Id3}:
\begin{equation}
\omega_L\left(\dot{a}_\eta, \pi_1\M\frac{1}{1-a_\eta}\otimes a_\delta \otimes\frac{1}{1-a_\eta}\right) = -\omega_L\left(\pi_1\M\frac{1}{1-a_\eta}\otimes \dot{a}_\eta \otimes\frac{1}{1-a_\eta}, a_\delta\right).
\end{equation}
For this purpose consider the identity
\begin{equation}
0= \langle \omega_L| \pi_2\M\frac{1}{1-a_\eta}\otimes \dot{a}_\eta\otimes \frac{1}{1-a_\eta}\otimes a_\delta\otimes \frac{1}{1-a_\eta},
\end{equation}
which vanishes because $\M$ is cyclic with respect to the large Hilbert space symplectic form. Replacing 
$\pi_2 = \inverttriangle\! \Big[\pi_1\otimes'\pi_1\Big]\triangle$ and acting with the coproduct gives
\begin{eqnarray}
0\lineup = \langle \omega_L| \inverttriangle\!(\pi_1\otimes'\pi_1)(\M\otimes'\mathbb{I}_{T\mathcal{H}}+\mathbb{I}_{T\mathcal{H}}\otimes'\M)
\Bigg[\frac{1}{1-a_\eta}\otimes'\frac{1}{1-a_\eta}\otimes \dot{a}_\eta\otimes \frac{1}{1-a_\eta}\otimes a_\delta\otimes \frac{1}{1-a_\eta}\nonumber\\
\lineup\ \ \ \ \ \ \ +\frac{1}{1-a_\eta}\otimes \dot{a}_\eta\otimes \frac{1}{1-a_\eta}\otimes'\frac{1}{1-a_\eta}\otimes a_\delta\otimes \frac{1}{1-a_\eta}+\frac{1}{1-a_\eta}\otimes \dot{a}_\eta\otimes \frac{1}{1-a_\eta}\otimes a_\delta\otimes \frac{1}{1-a_\eta}\otimes'\frac{1}{1-a_\eta}\Bigg].\ \ \ \ \ \ \ \ 
\end{eqnarray}
Some cross terms drop out since $\pi_1$ acts on the tensor product of two or more states. What is left is
\begin{eqnarray}
0\lineup =\langle \omega_L|\left(\pi_1\M \frac{1}{1-a_\eta}\otimes \dot{a}_\eta\otimes \frac{1}{1-a_\eta}\right)\otimes a_\delta\nonumber\\ 
\lineup\ \ \ +\langle \omega_L| \left(\pi_1 \M \frac{1}{1-a_\eta}\otimes \dot{a}_\eta\otimes \frac{1}{1-a_\eta}\otimes a_\delta\otimes \frac{1}{1-a_\eta}\right)\otimes a_\eta\nonumber\\
\lineup\ \ \ +\langle \omega_L| a_\eta\otimes\left(\pi_1\M \frac{1}{1-a_\eta}\otimes \dot{a}_\eta\otimes \frac{1}{1-a_\eta}\otimes a_\delta\otimes \frac{1}{1-a_\eta}\right)\nonumber\\
\lineup\ \ \ +\langle \omega_L| \dot{a}_\eta\otimes \left(\pi_1\M \frac{1}{1-a_\eta}\otimes a_\delta\otimes \frac{1}{1-a_\eta}\right).
\end{eqnarray}
The second and third terms cancel out by antisymmetry of the symplectic form, while the first and last terms reproduce \eq{Id3}.

\end{appendix}

\end{document}